\documentclass[a4paper,11pt]{article}

\usepackage{jcappub} % for details on the use of the package, please
\usepackage{mathptmx}       % selects Times Roman as basic font
\usepackage{helvet}         % selects Helvetica as sans-serif font
\usepackage{courier}        % selects Courier as typewriter font
\usepackage{type1cm}        % activate if the above 3 fonts are
\usepackage{makeidx}         % allows index generation
\usepackage{graphicx}        % standard LaTeX graphics tool
\usepackage{mathtools}
\usepackage{multirow}
\usepackage{multicol}        % used for the two-column index
\usepackage[bottom]{footmisc}% places footnotes at page bottom

\def\apj{The Astrophysical Journal}

\def\mnras{Mon. Not. Roy. Astron. Soc.}

\def\apjs{Astrophys. J. Suppl.}
\def\aap{Astron. Astrophys.}

\def\jcap{Journal of Cosmology and Astroparticle Physics}
\def\prd{Phys. Rev. D}

\title{Observational constraints on Quintessence models of dark energy}
\author[a]{Archana Sangwan}
\author[a,b]{Ashutosh Tripathi}
\author[a]{H. K. Jassal}

\affiliation[a]{Indian Institute of Science Education and Research
Mohali,\\ SAS Nagar, Mohali 140306, Punjab, India.}
\affiliation[b]{Center for Field Theory and Particle Physics and 
Department of Physics,\\ Fudan University, 200433 Shanghai, China.}

\emailAdd{archanakumari@iisermohali.ac.in}
\emailAdd{ashutosh\_tripathi@fudan.edu.cn}
\emailAdd{hkjassal@iisermohali.ac.in}

\abstract{
  Scalar fields aptly describe equation of state of dark
energy.
The scalar field  models were initially proposed to
circumvent the fine tuning problem of cosmological constant.
However, the model parameters also need a fine tuning of their own and
it is important to use  observations to determine these
parameters.
In this paper, we use a combination of low redshift data to constrain
the of canonical scalar field parameters.
For this analysis, we use the Supernova Type Ia Observations, the
Baryon Acoustic Observations and the Hubble parameter measurement data.
We consider scalar field models of the thawing type of two different
functional forms of potentials.
The constraints on the model parameters are more stringent than those from earlier
observations although these datasets do not rule out the models
entirely.
The parameters which let dark energy dynamics closely emulate 
that of a cosmological constant are preferred. 
The constraints on the parameters are suitable  priors for further quintessence dark energy studies.  
 
}

\begin{document}
\maketitle
\flushbottom

\section{Introduction}\label{sec::intro}
The discovery of late time cosmic acceleration by the Supernova Type
Ia observations has been one of the most important results in
cosmology  \cite{riess:1998cb,perlmutter:1998np}.
%These observations confirmed  earlier reports that the universe may
%be dominated by a cosmological constant \cite{Zehavi:1999fm,Carroll:1991mt,cc2}.
Nearly two-thirds of the energy  of the universe is  due to the
cosmological constant or an alternative description 
called the dark energy\cite{caldwell2004}.  
The presence of dark energy has been further confirmed by many different
observations such as Baryon  Acoustic Oscillations \cite{bao1} and
Cosmic Microwave Background observations \cite{cmb1,cmb2}.
More recently, data from direct measurements of the Hubble parameter
has also shown to be a useful probe of low redshift evolution of the
universe \cite{hz1,hz2,hz3}.

Dark energy can be modeled by invoking the presence of a
cosmological constant model  for which  the  value of the
equation of state parameter is $w=-1$ \cite{cc1,cc2} and this model is
consistent with observational data \cite{cc3,cc4_obs,cc5_obs}. 
Einstein's cosmological constant $\Lambda$ is attributed to the
zero-point energy of the vacuum, with a constant energy density
$\rho$, a negative pressure $P$ with an equation of state given by $w
\equiv  P/\rho =-1$. 
Observations do, however, allow a $w$ which is different from that of
a cosmological constant and has a dynamical nature i.e. it varies with
time.
The variation with time is achieved by extending the description of
barotropic fluid equation of state parameter to be a function of time or
the scale factor.
A few parameterisations which have been proposed are described in 
\cite{fluid,reconst} and there are non baroptropic fluids such as the
Chaplygin gas in \cite{chap1}.

A slowly varying scalar field have been proposed to be a viable
substitute for the cosmological constant as the negative equation of
state parameters arises naturally in these scenarios.
The scalar field  models include those based on canonical scalar field
matter such as the quintessence
\cite{quin,quin2,quin3,quin4,quin5,quin6,quin7}, kinetic energy driven
k-essence \cite{kess1,kess2} and others like tachyon
\cite{Padmanabhan:2002cp,Bagla:2002yn}.
There is at present no consensus as to which of these models better
describe dark energy.
Although proposed to do away with the fine tuning problem of the
cosmological constant, the scalar field models have fine tuning 
requirements of their own.
The potential parameters need to be fine tuned such that the
acceleration of the universe begins after a sufficiently long matter
dominated era in order that the  large scale structures form. 
Among the models listed above, the quintessence model is described by
a canonical scalar field.
For a slowly varying field, the scalar field potential model
universe has a positive acceleration.

Since a  large amount of data is now available and there is a large variety of
observations, it is  possible to constrain cosmological parameters to
to better precision than before.
This is especially true of observations which have constraints
orthogonal to each other and hence the combined range is much smaller
than that allowed  by individual datasets.
Many observations such as supernovae (SNIa) data
\cite{riess:1998cb,perlmutter:1998np,snia3,snia4,snia5,snia6,snia7,snia8,snia9},
Baryonic Acoustic oscillation (BAO) data
\cite{bao1,bao2,bao3,bao4,bao5,bao6,bao7} and Hubble distance (H(z))
measurements using different methods by \cite{bao3,bao4,moresco2012,moresco2015,moresco2016,Zhang2012,Simon:2004tf,Chuang2012b,hz1}, 
compiled in \cite{hz1,hz2,hz3} can be used for constraining models.

In this work, we revisit the quintessence dynamics in the light of
more recent and diverse cosmological observations.
We restrict ourselves to low redshift, distance measurement data.
The main motivation in this work is to study the present constraints with a view to reduce the priors.
We consider different quintessence scenarios, with 
different scalar field potentials.
These different scenarios have
been broadly classified as thawing and freezing type
\cite{review,Pantazis,linder2008,caldwell2005,huterer2006,linder2006}. 
This broad classification is based on the whether the equation of
state parameters is cosmological constant like in the past, or if this
behaviour is at later time.   
While it may be expected that the equation of state parameter can be
effectively constrained by assuming dark energy to be a fluid, it is
important to explicitly study different scalar field models.
The equation of state parameter depends on the time evolution of the
scalar field and the functional form of the scalar field potential.
In this work, we determine constraints on the equation of state parameters for
different thawing scalar field models.
Earlier similarly motivated studies include
\cite{watson2003,scherrer2007,dutta2011,chang2016,steinhardt1999,Macorra2000,Nunes2001,corasaniti2003,Slepian:2013ug,Akrami:2017cir,Casas:2017wjh,Ryan:2018aif,Avsajanishvili:2017zoj}.
Structure formation in dark energy scenarios has also been studied in \cite{Unnikrishnan:2008qe,Jassal:2009ya,pert1,Jassal:2009gc,Rajvanshi:2018xhf,Nazari-Pooya:2016bra}.
Recent study shows that the scalar fields fir observations better than
the  $\Lambda$CDM model, but the difference is not significant
\cite{Ooba:2018dzf}.

This paper is structured as follows.
After introduction in section \ref{sec::intro}, in section \ref{sec::quin}, we 
discuss cosmological  equations for quintessence scalar field model.
In section \ref{sec::sol_to_coseq}, we show the solutions of the equations for
different potentials.
The key results are discussed in section \ref{sec::results} and we
summarize and  conclude in section \ref{sec::conclusions}.

\begin{figure*}[t]
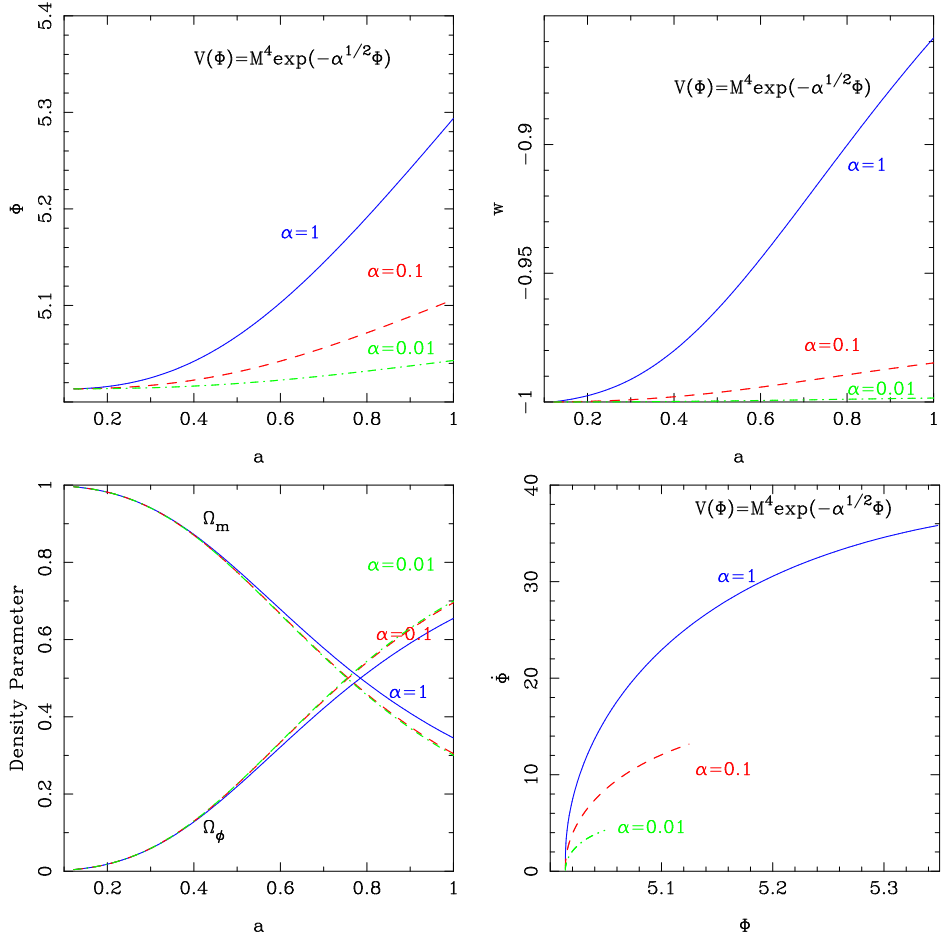

\centering
\begin{tabular}{cc}
\includegraphics[scale=0.4]{fig1_a.ps}&\includegraphics[scale=0.4]{fig1_b.ps}\\ 
\includegraphics[scale=0.4]{fig1_c.ps}&\includegraphics[scale=0.4]{fig1_d.ps}
\end{tabular}
\caption{The plots in the figure represent evolution of different quantities 
  corresponding to  the  potential $ V(\phi) = M^{4} exp(-\alpha^{1/2} \phi /
  M_{p})$. We consider three different values of  
 $\alpha =0.01,~0.1$ and $~1$ and we  have rescaled the variable $\phi$ as $\Phi=\phi/M_p$. 
 In the first row, the plot on the left  shows the variation of $\Phi$
 as a function  of scale factor for different values of $\alpha$. The
 plot on the right   shows the behavior of equation of state parameter $w$ 
 as the scale factor varies. In the second row, the
  plot on the left is for  
  energy density parameter of the field and matter and the   figure
  on the right  is the phase plot  for the exponential 
  potential. Here, the solid curve represents results for $\alpha$=1,
 the dashed curve is drawn for $\alpha$=0.1 and the dot-dashed curve is for $\alpha$=0.01.}  
\label{fig::thaw_exp_the}
\end{figure*}

\begin{table}
\begin{tabular}{|c|c|c|c|c|l|}
\cline{1-4}
Potential&Parameter & Lower Limit & Upper Limit\\ \hline
\multicolumn{1}{ |c  }{\multirow{3}{*}{$V(\phi) = M^{4} e^{(-\alpha^{1/2} \phi / M_{p})}$} } &
\multicolumn{1}{ |c| }{$\Omega_m$} & 0.01 & 0.6     \\ \cline{2-4}
\multicolumn{1}{ |c  }{}                        &
\multicolumn{1}{ |c| }{$w_0$} & -1.0 & 1.0     \\ \cline{2-4}
\multicolumn{1}{ |c  }{}                        &
\multicolumn{1}{ |c| }{$\alpha$} & 0.01 & 190.0     \\ \cline{1-4}
\multicolumn{1}{ |c  }{\multirow{2}{*}{$V(\phi) = M^{4-n}  \phi ^n$} } &
\multicolumn{1}{ |c| }{$\Omega_m$} & 0.01 & 0.6    \\ \cline{2-4}
\multicolumn{1}{ |c  }{}                        &
\multicolumn{1}{ |c| }{$w_0$} & -1.0 & 1.0     \\ \cline{2-4}
\multicolumn{1}{ |c  }{}                        &
\multicolumn{1}{ |c| }{$\phi_{in}$} & 1.0 & 10.0     \\ \cline{1-4}
\end{tabular}
\caption{This table lists the priors used for parameter fitting in case of both the potentials.}
\label{table::priors}
\end{table}

\section{Quintessence Dynamics}
\label{sec::quin}
We consider a canonical scalar field $\phi$ minimally coupled, i.e. experiencing
gravity passively through the spacetime curvature and a
self-interaction described by the scalar field potential
V($\phi$) and with a canonical  kinetic energy contribution. 
The action for a quintessence field  is therefore given by 
\begin{equation}
S = \int d^4x\sqrt{-g}\begin{pmatrix}-\frac{1}{2}(\nabla\phi)^2 - V(\phi)\end{pmatrix}
\end{equation}
where
\begin{equation}
(\nabla\phi)^2 = g^{\nu\mu}\partial_{\nu}\phi\partial_{\mu}\phi
\end{equation}
In a flat Friedmann background, the pressure and energy density of 
a homogeneous scalar field are given by  
\begin{eqnarray} \label{pandrho}
P=\frac{\dot{\phi}^2}{2} - V(\phi) \\ \nonumber          
\rho=\frac{\dot{\phi}^2}{2} + V(\phi).
\end{eqnarray}
The equation of state, which  in general is time varying, is defined
as 
\begin{equation}
w=\frac{P}{\rho}.
\end{equation}
The equation of motion for the scalar field, the Klein-Gordon equation
takes the form
\begin{equation}\label{kleingordon}
\ddot{\phi} + 3H\dot{\phi} = - \frac{dV}{d\phi},
\end{equation}
follows from functional variation of the Lagrangian and is
interchangeable with the continuity equation.

\begin{figure*}[t]
\centering
\begin{tabular}{ccc}
\includegraphics[scale=0.33,angle=270]{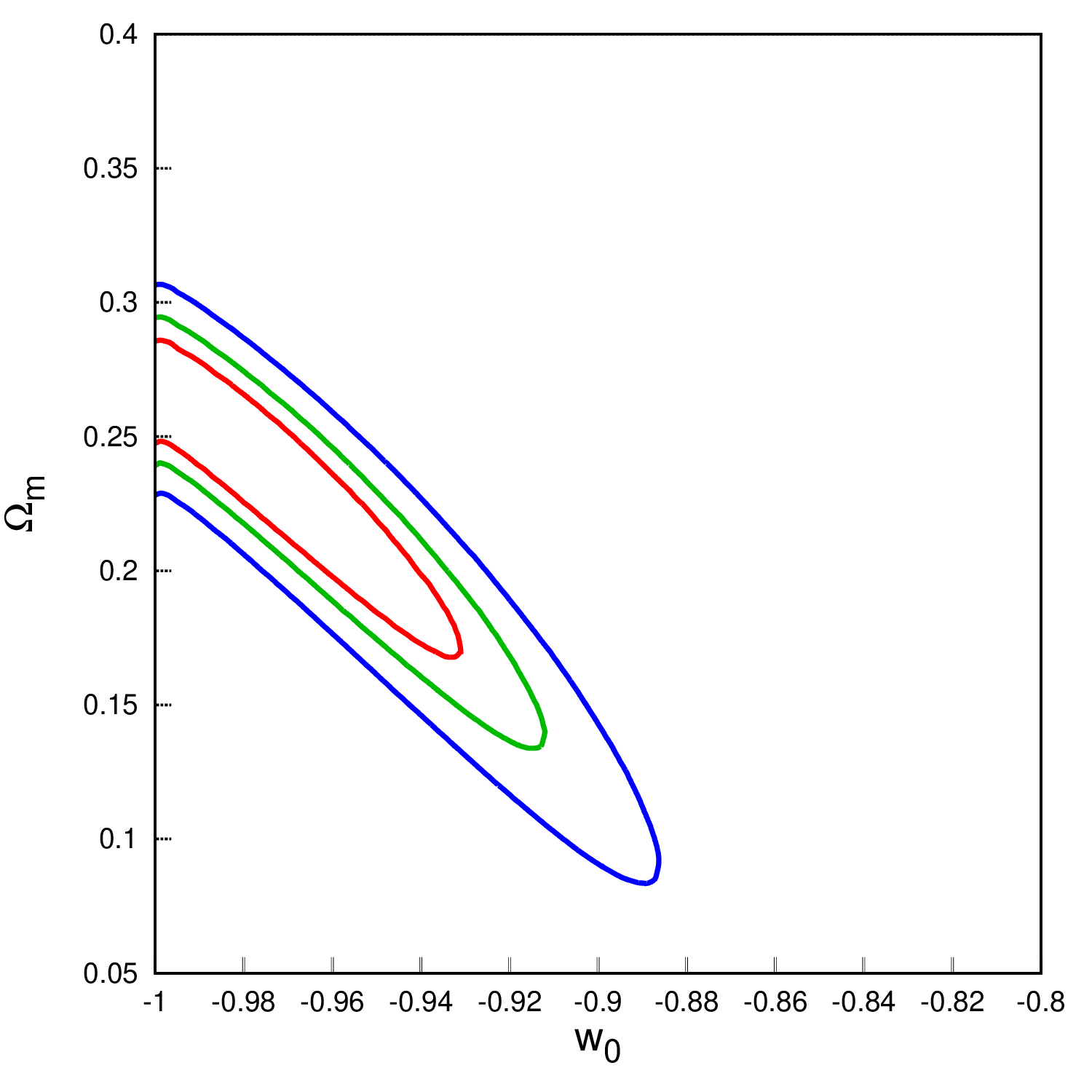}&\includegraphics[scale=0.33,angle=270]{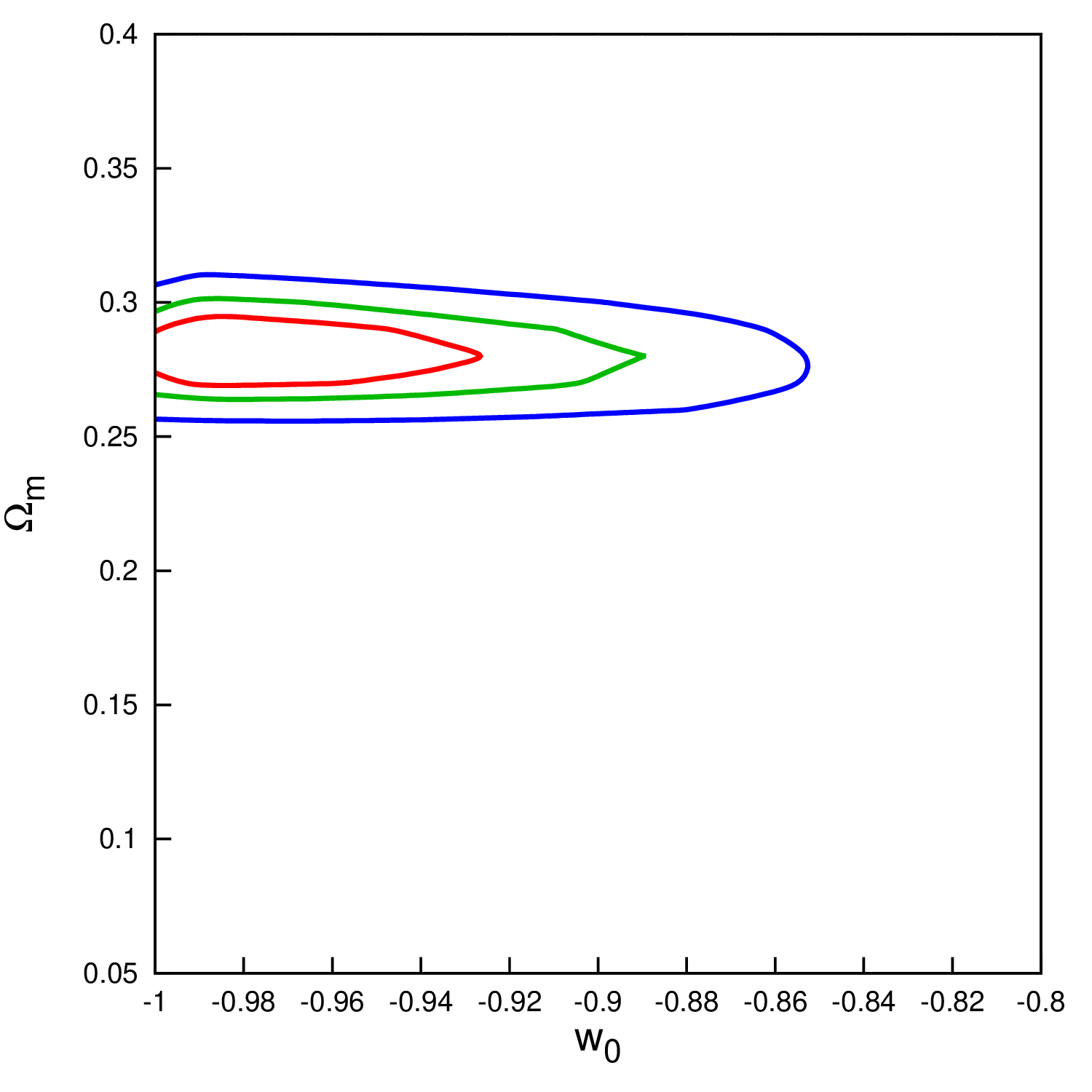}&\includegraphics[scale=0.33,angle=270]{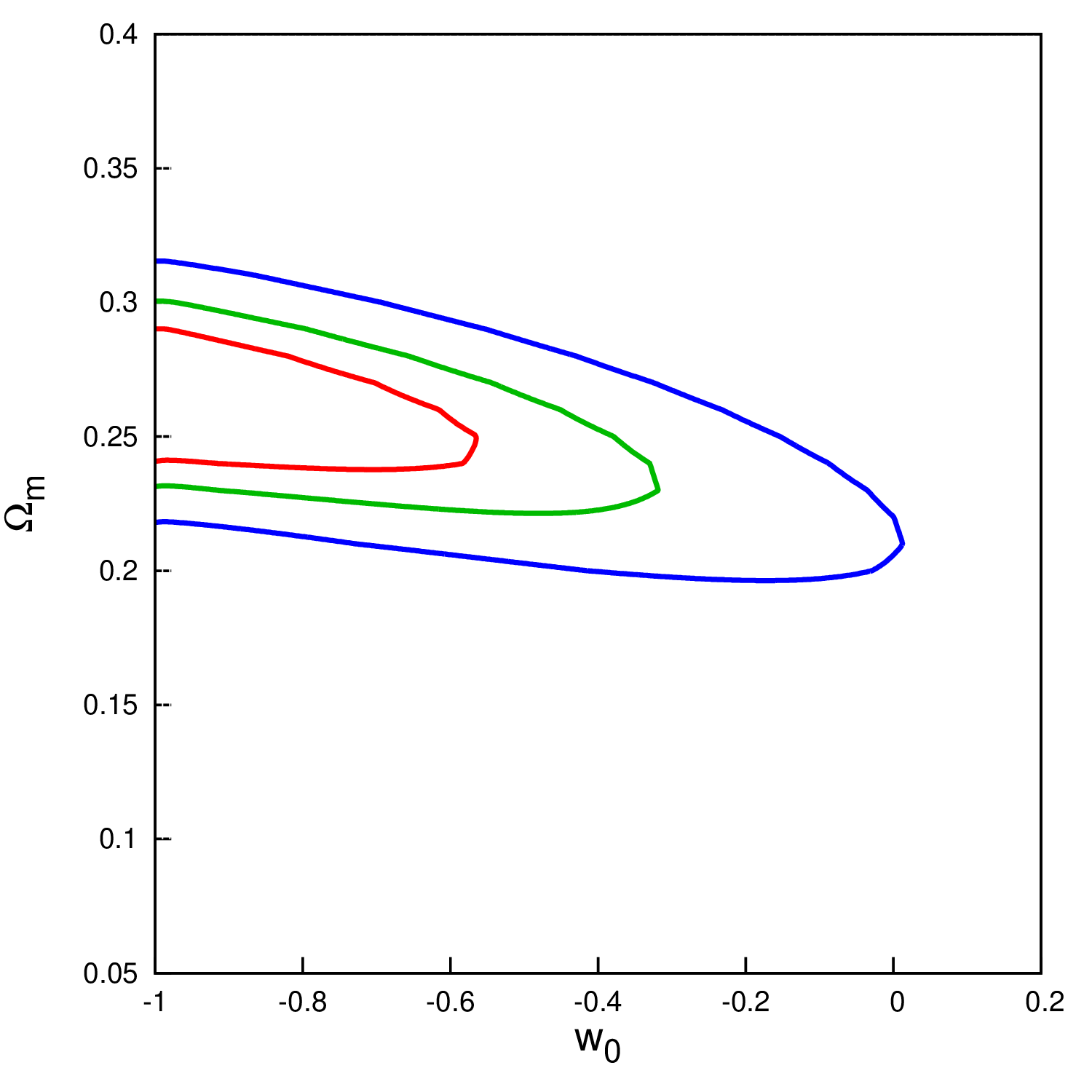}\\  
\end{tabular}
\caption{The figure represents $1 \sigma$, $2 \sigma$ and 
  $3 \sigma$ confidence  contours in $w_0-\Omega_m$ plane for the  thawing
  potential $V = M^4\exp(-\sqrt\alpha\Phi)$. From left to right,
  the plot are  for the  SNIa, BAO and H(z)
  data respectively. Here, we have marginalised over the parameter $\alpha$.}  
\label{fig::thaw_exp_data}
\end{figure*}

\begin{table}
\begin{tabular}{|l|l|l|l|} \hline
  Data set&$3\sigma$ confidence&$\chi_m^2$&Best Fit Model \\
  \hline

\multicolumn{4}{|c|}{$V = M^4\exp(-\sqrt\alpha\Phi)$} \\
\hline
SNIa& -1.0$\leq$  $w_0$ $\leq$-0.63&&$w_0$=-0.97\\ 
&0.01$\leq$  $\Omega_m$ $\leq$0.31&563.42&$\Omega_m$=0.24\\
& 0.01$\leq$ $\alpha$ $\leq$1.0&&$\alpha$=0.03\\ 
&&&\\ \hline
BAO& -1.0$\leq$  $w_0$ $\leq$-0.85&&$w_0$=-0.99\\ 
& 0.26$\leq$  $\Omega_m$ $\leq$0.31&2.35&$\Omega_m$=0.28\\
& 0.01$\leq$ $\alpha$ $\leq$ 1.0&&$\alpha$=0.07\\
&&&\\ \hline
H(z)& -1.0$\leq$  $w_0$ $\leq$0.14&&$w_0$=-1.0\\ 
& 0.19$\leq$  $\Omega_m$ $\leq$0.32&17.04&$\Omega_m$=0.26\\
& 0.01$\leq$  $\alpha$ $\leq$1.0&&$\alpha$=1.0\\
&&&\\ \hline
\end{tabular}
\caption{The above table shows the 3$\sigma$ confidence limit for all the three data for the potential $V = M^4\exp(-\sqrt\alpha\Phi)$ and the value of the parameters corresponding to minimum value of $\chi^2$ for the exponential potential.}
\label{tab:thaw_exp_chi}
\end{table}

%\begin{figure*}[t]
%\centering
%\begin{tabular}{ccc}
%\includegraphics[scale=0.33,angle=270]{fig3_a.ps}&\includegraphics[scale=0.33,angle=270]{fig3_b.ps}&\includegraphics[scale=0.33,angle=270]{fig3_c.ps}\\  
%\end{tabular}
%\caption{The figure represents $1 \sigma$, $2 \sigma$ and 
%  $3 \sigma$ confidence  contours in $\sqrt{\alpha}\Phi-V_0$ plane for
%  the exponential
%  potential $V = M^4\exp(-\sqrt\alpha\Phi)$. The plots are arranged using the same scheme as in figure~\ref{fig::thaw_exp_data}. Here, $V_0$ is scaled by the square of the present value of Hubble parameter ($H_0^2$).}  
%\label{fig::thawexp_v0_alphaphi}
%\end{figure*}

\begin{figure*}[t]
\centering
\begin{tabular}{ccc}
\includegraphics[scale=0.33,angle=270]{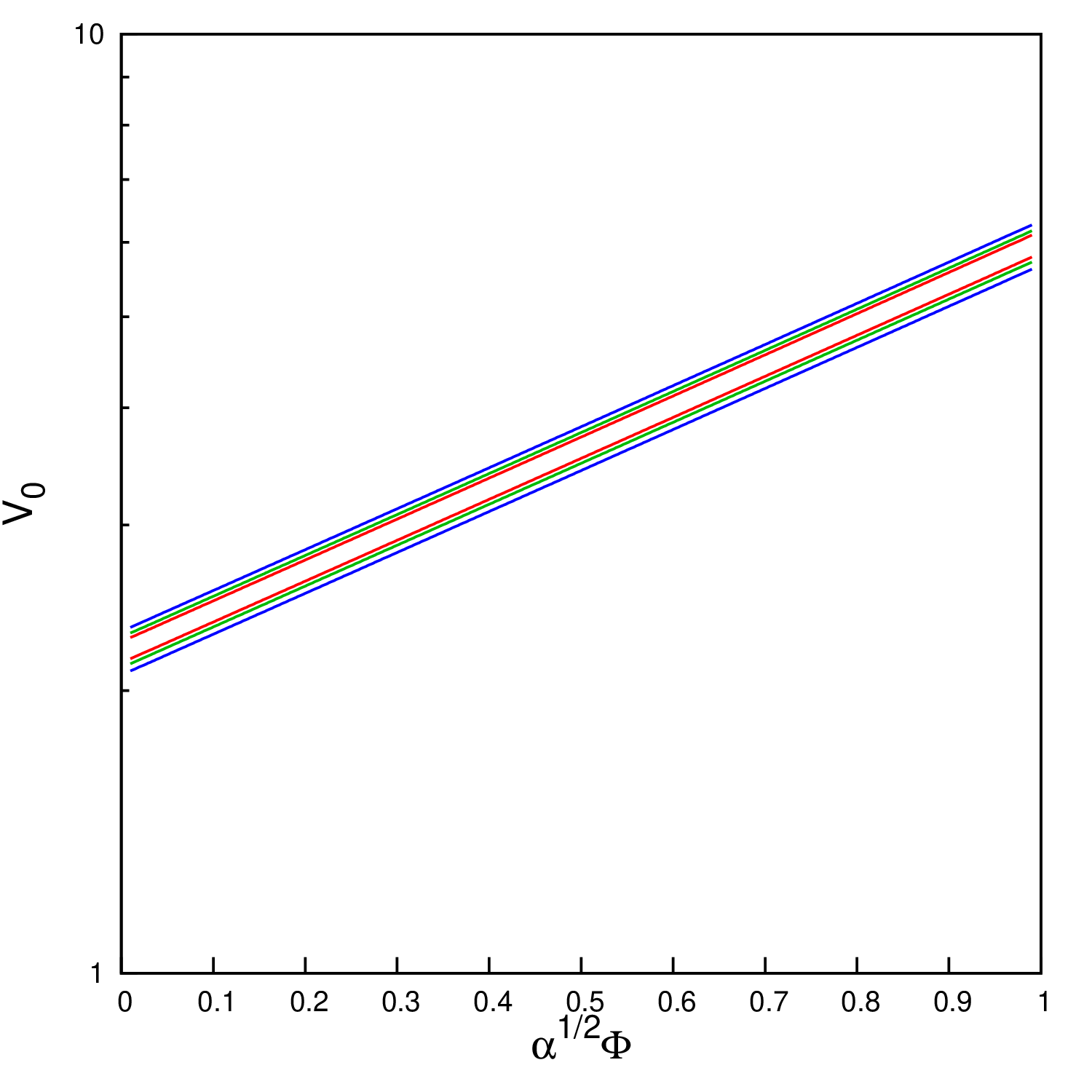}&\includegraphics[scale=0.33,angle=270]{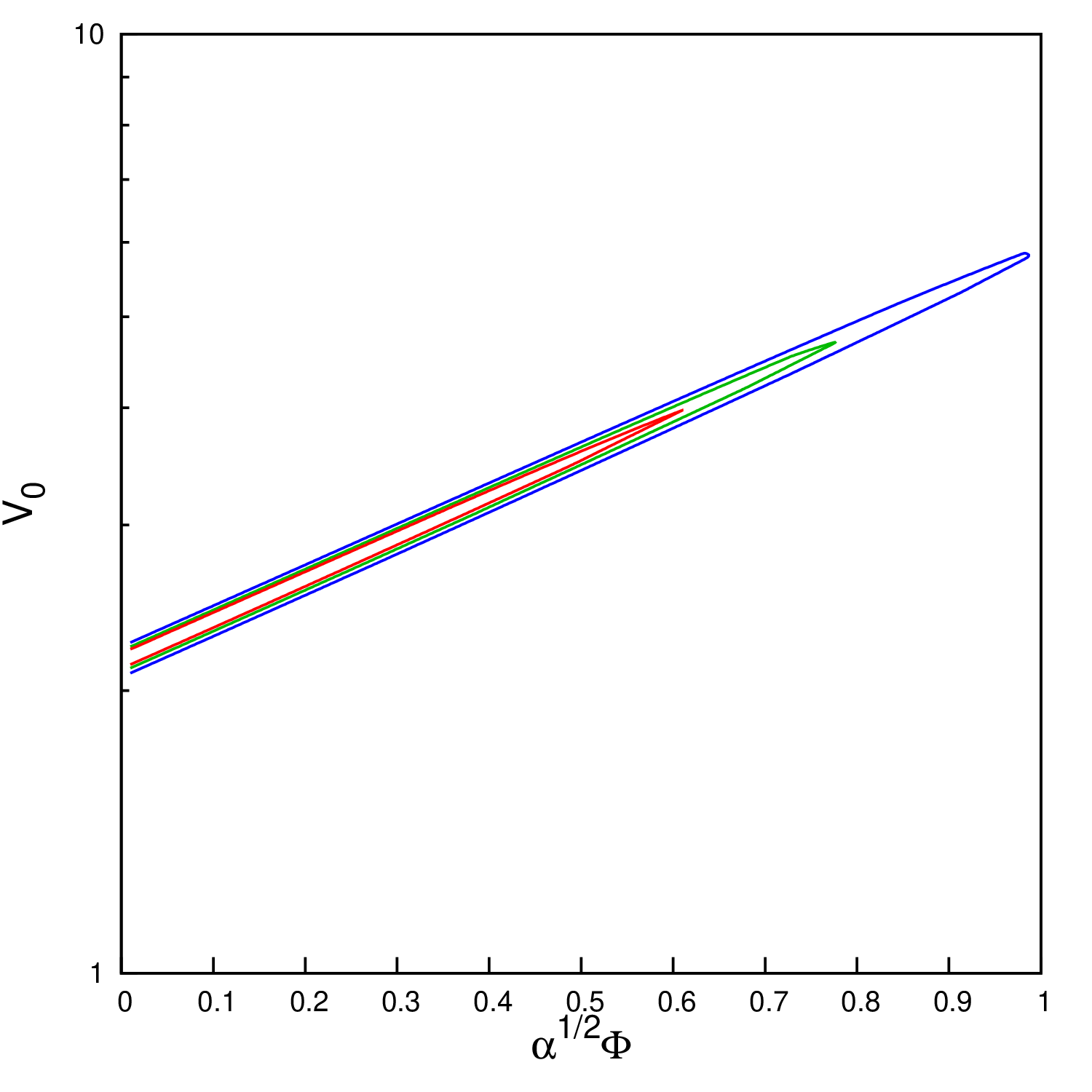}&\includegraphics[scale=0.33,angle=270]{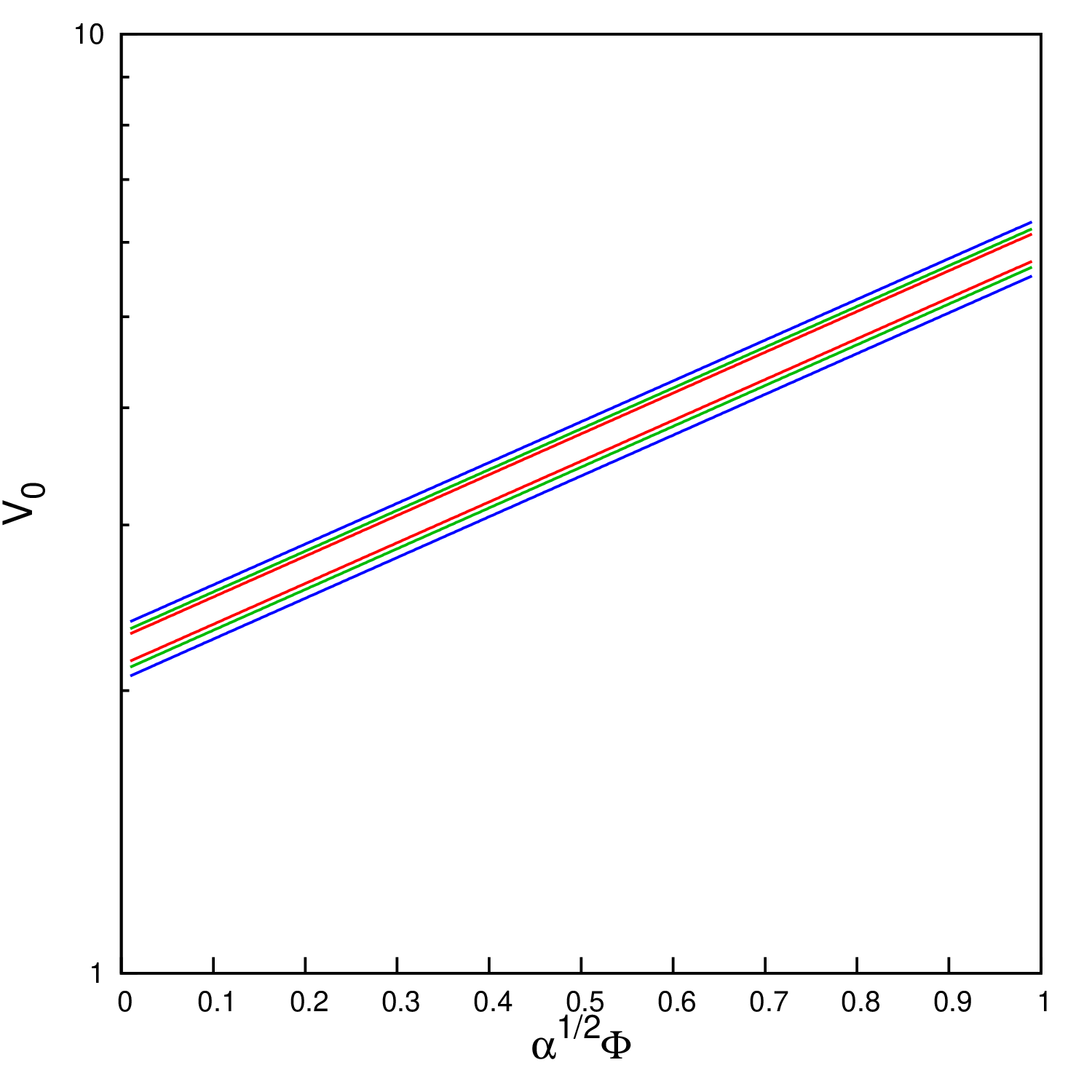}\\
\includegraphics[scale=0.33,angle=270]{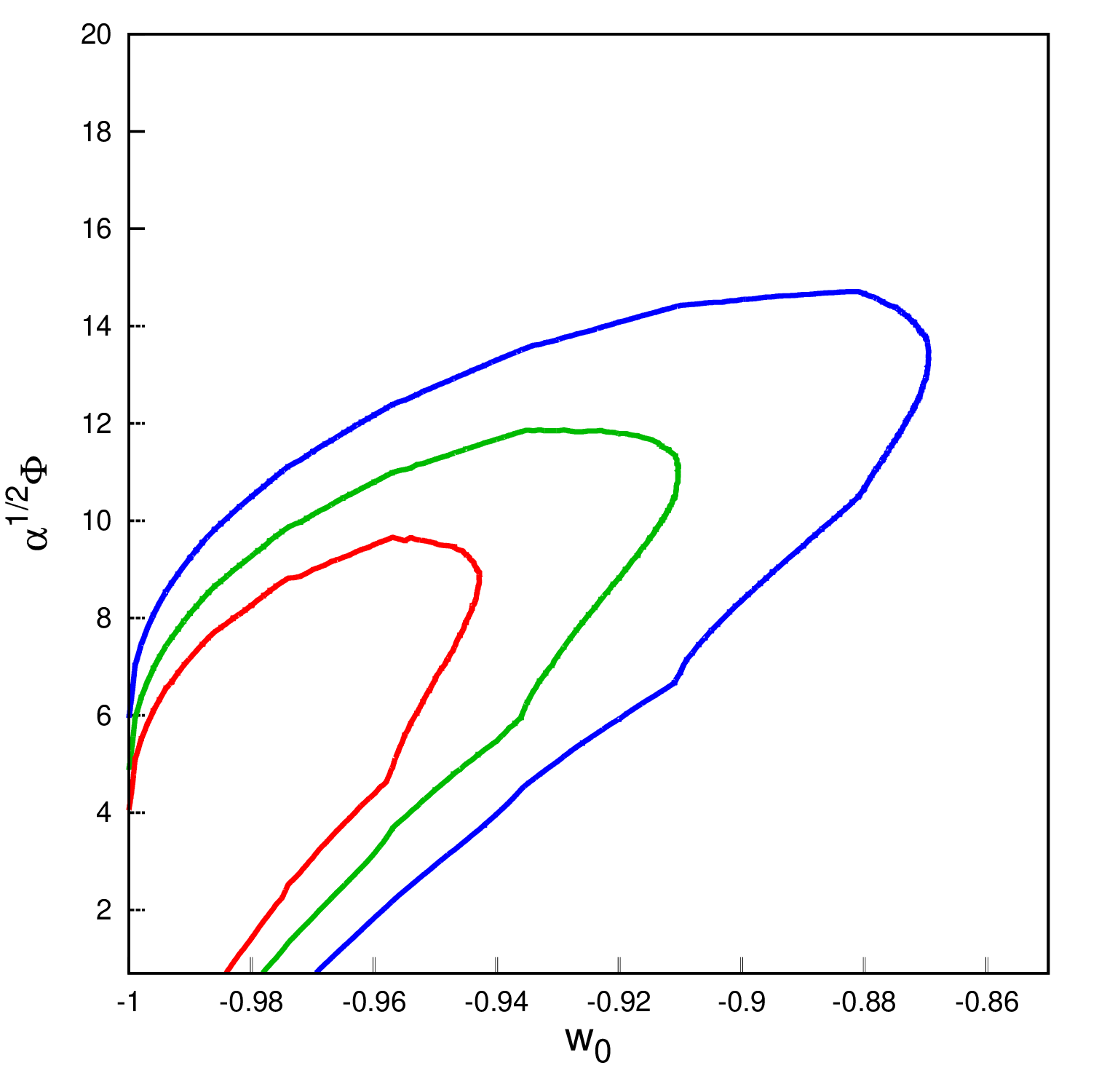}&\includegraphics[scale=0.33,angle=-90]{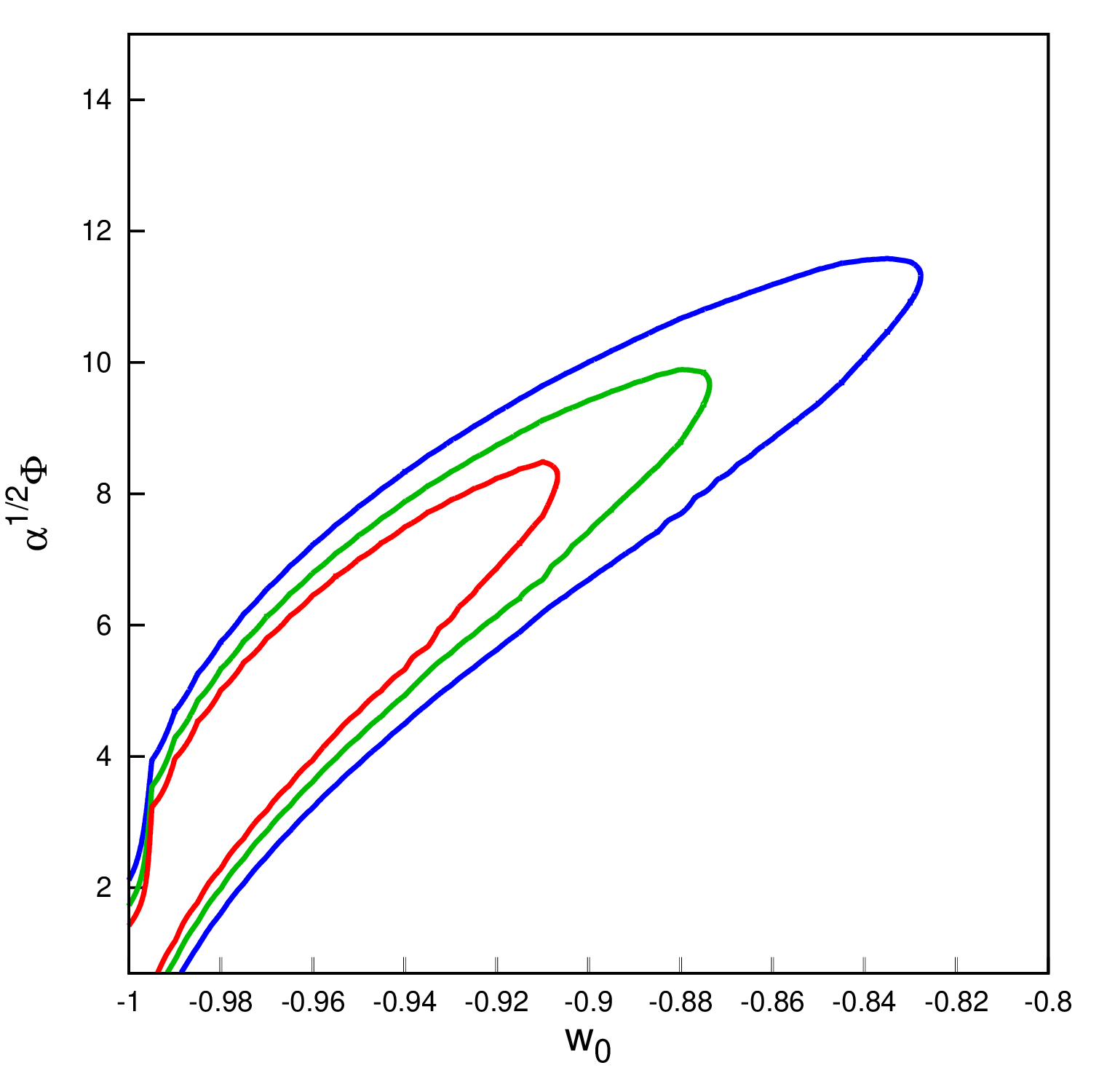}&\includegraphics[scale=0.33,angle=-90]{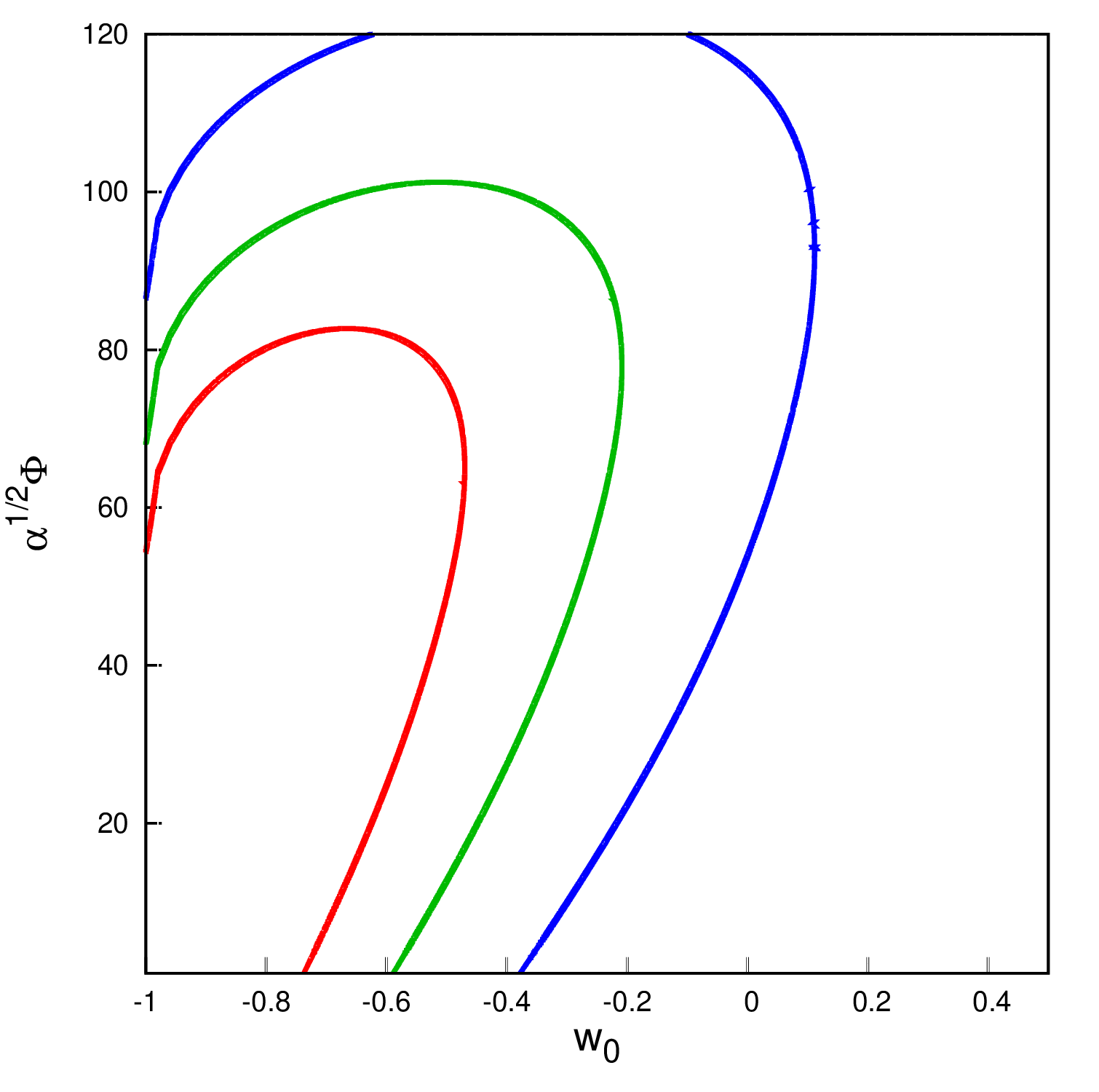}\\  
\includegraphics[scale=0.33,angle=-90]{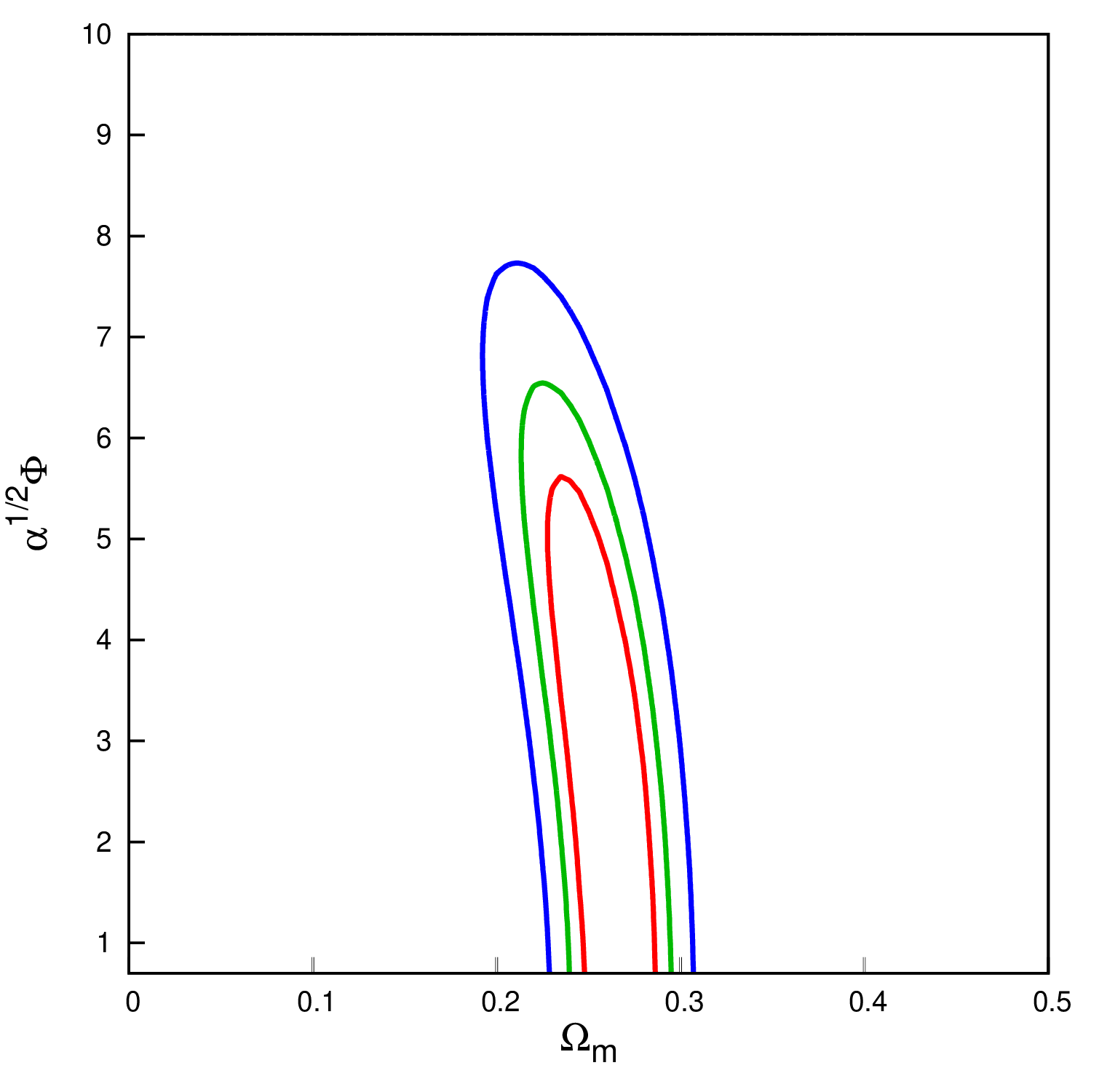}&\includegraphics[scale=0.33,angle=-90]{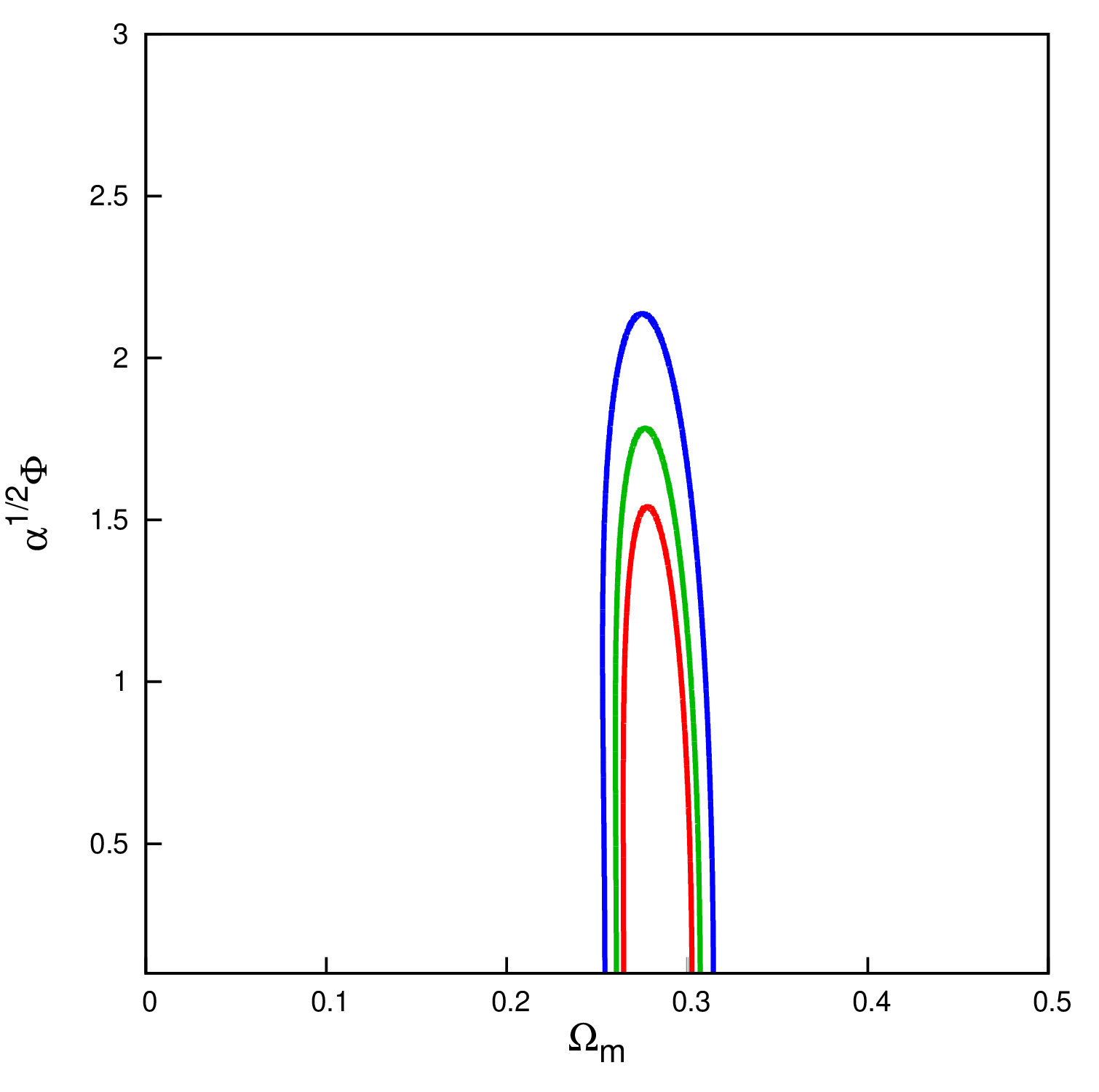}&\includegraphics[scale=0.33,angle=-90]{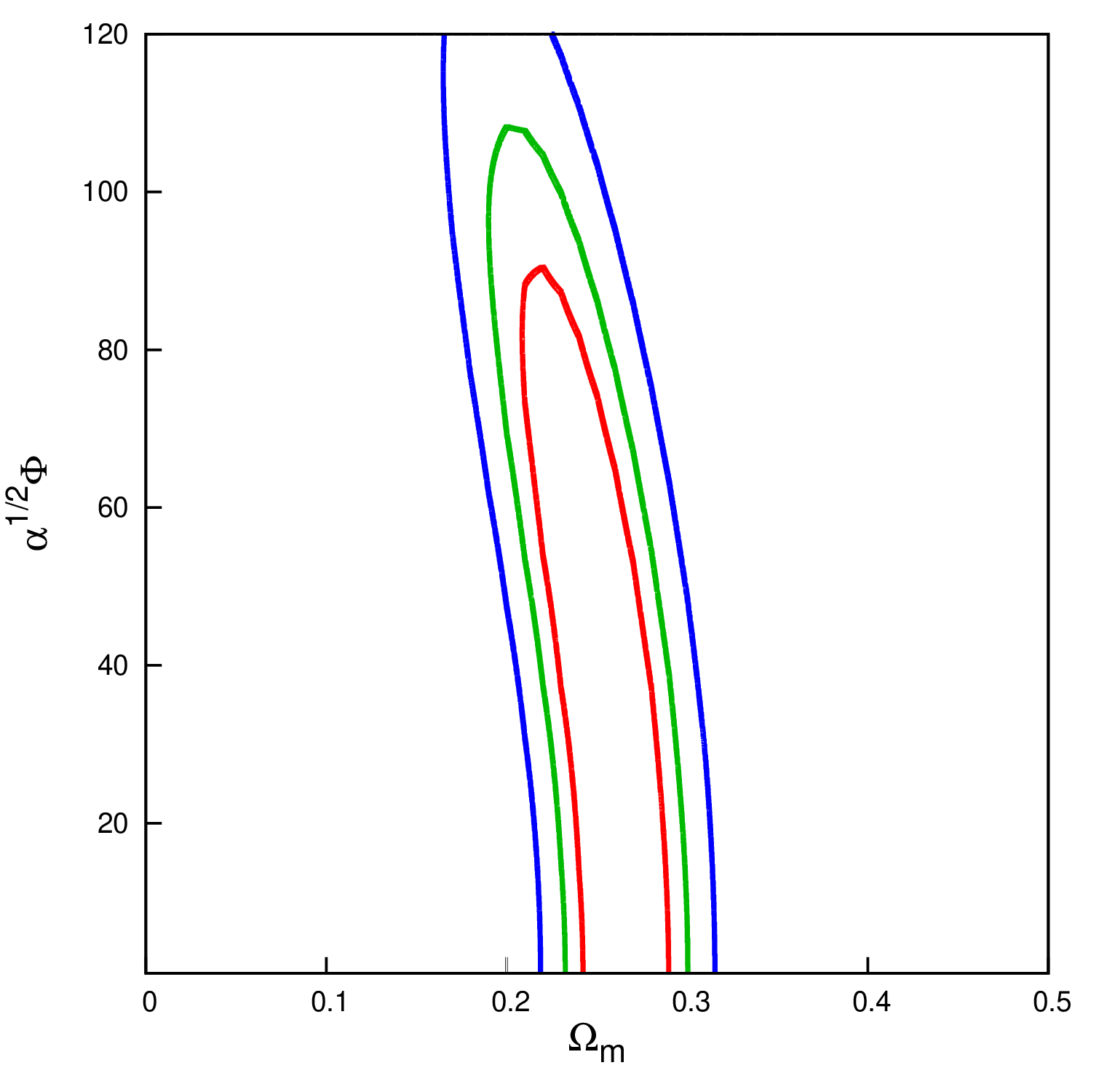}\\  
\end{tabular}
\caption{The figure represents $1 \sigma$, $2 \sigma$ and $3 \sigma$ confidence  contours in $\sqrt{\alpha}\Phi-V_0$ plane for the exponential
 potential $V = M^4\exp(-\sqrt\alpha\Phi)$ in the first row.
Here, $V_0$ is scaled by the square of the present value of Hubble parameter ($H_0^2$).
The plots in second row represents confidence  contours in $w_0-\sqrt{\alpha}\Phi$ plane and contours in plane $\Omega_m-\sqrt{\alpha}\Phi$ are given in third row. The scheme of the plots is same as in figure \ref{fig::thaw_exp_data}.}  
\label{fig::thawexp_alphaphi}
\end{figure*}

The evolution of a spatially flat universe is described by Friedmann equations,
\begin{equation}\label{friedmann1}
H^2 = \left( \frac{\dot{a}}{a}\right) ^2 = \frac{8 \pi G \rho}{3}
\end{equation}
\begin{equation}\label{friedmann2}
\frac{\ddot{a}}{a} = -\frac{4 \pi G}{3} (\rho + 3P)
\end{equation}
where $H$ is the Hubble parameter, $\rho$ and $P$ denote the total
energy density and pressure of all the components present in the
universe at a given epoch.
Using equation (\ref{pandrho}) yields 
\begin{equation}\label{field1}
H^2 = \frac{8 \pi G}{3} \left[  \frac{1}{2} \dot{\phi}^2 + V(\phi) \right] ,
\end{equation} 
\begin{equation}\label{field2}
\frac{\ddot{a}}{a} = -\frac{8 \pi G}{3} \left[  \dot{\phi}^2 - V(\phi) \right] .
\end{equation}
For an accelerating universe $\dot{\phi}^2 < V(\phi)$.
This implies that one requires an almost flat potential for an
accelerated expansion.
The equation of state for the scalar field $\phi$ is given by
\begin{equation}\label{wphi}
w = \frac{ \dot{\phi}^2 - 2V(\phi)}{\dot{\phi}^2 + 2V(\phi)}.
\end{equation}

Depending on the evolution of $w$, different quintessence models
are classified into two broad categories \cite{scherrer2007,dutta2011,steinhardt1999,Gupta2014,Chiba2009,scherrer2005,schimd2006,Sahlen2006,Chiba2005}. 
The first  corresponds to thawing models, in which the field is
nearly frozen by a Hubble damping during the early cosmological epoch
and it starts to evolve at late times. 
Here the field is displaced from its frozen value recently, when it
starts to roll down to the minimum. 
In this case, the evolution of $w$ is characterised by the growth from
$-1$, at early times the equation of state is $w\approx$ -1, but grows
less negative with time. 
We analyse the following concave potentials for thawing behaviour \cite{Ferreira1997,Ferreira1997b,Kallosh2003,Linde1991,Linde1994}.
\begin{itemize}
\item Exponential potential \cite{quin5,Halliwell:1986ja,Barreiro:1999zs,Rubano:2001su,Heard:2002dr}:
\begin{equation}    
V = M^{4}\exp(-\sqrt{\alpha} \phi / M_{p})
\end{equation}
\item Polynomial (concave) potential : 
\begin{equation}
 V = M^{4-n} \phi^n.
\end{equation}
\end{itemize}
For the potential described by a polynomial, we consider $n=1, 2, 3$.
These different values correspond to potentials with different shapes.
The other class of potentials consists of a field which was already
rolling towards minimum of its potential, prior to the onset 
of acceleration, but  slows down because of the shallowness of the
potential at late times and comes to a halt as it begins to dominate
the universe. 
For freezing models, the equation of state parameter $w$ approaches $-1$.
For this work, we will focus on homogeneous scalar fields belonging to
thawing class.

\section{Solutions to cosmological equations}
\label{sec::sol_to_coseq}
\begin{figure*}[t]
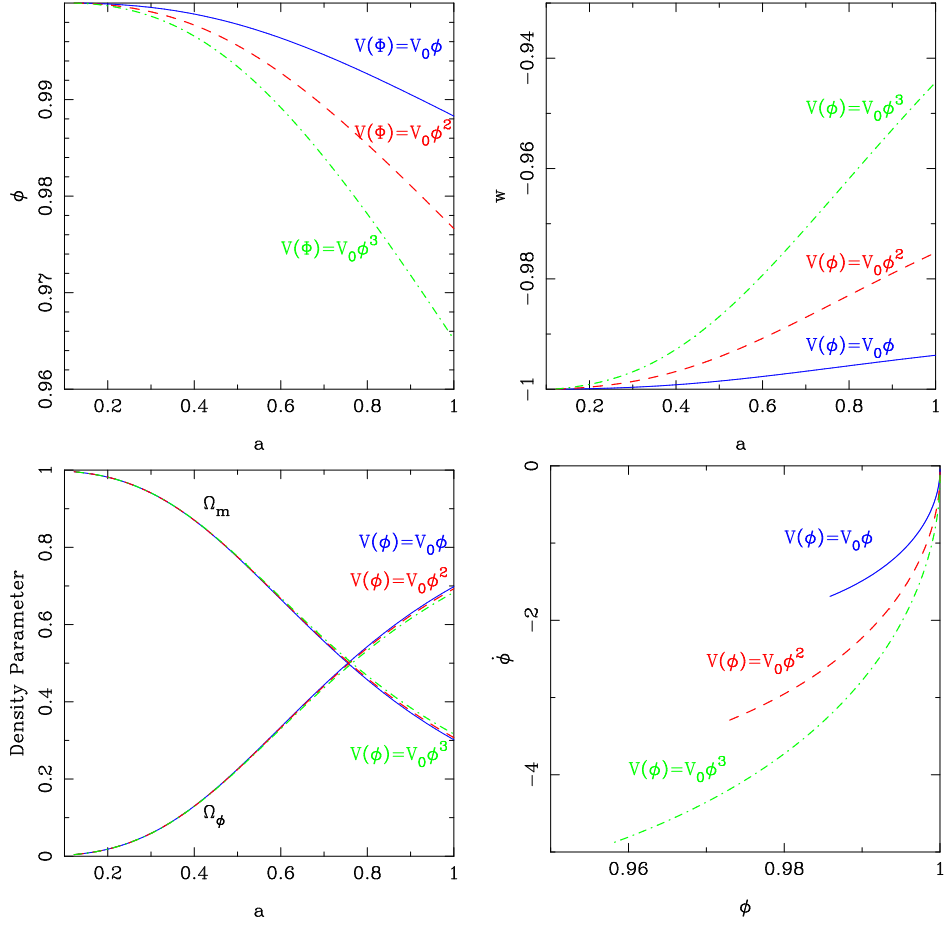

\centering
\begin{tabular}{cc}
\includegraphics[scale=0.4]{fig5_a.ps}&\includegraphics[scale=0.4]{fig5_b.ps}\\ 
\includegraphics[scale=0.4]{fig5_c.ps}&\includegraphics[scale=0.4]{fig5_d.ps}
\end{tabular}
\caption{The plots in this figure represent the theoretical results obtained
  for potential $ V(\phi) = M^{4-n} \phi^n$, for $n$ =1, 2 and 3. In this figure,
 the solid line represents results for $n=1$, dashed curve is drawn for quadratic
 potential, $n=2$ and dotted-dashed curve is for cubic power potential, $n=3$.
  In the first row, the plot on the left shows the variation of $\phi$ as a function
  of scale factor. The figure on the right shows the evolution of equation of state parameter $w$
  for the potential as scale factor changes. In the second row, the
  plot on the left  is for variation of 
  energy density parameter for the field ($\Omega_{\phi}$) and matter ($\Omega_m$) as a function of scale factor and the
  figure on the right is
  the phase plot  for the power potential.} 
\label{fig::thaw_the}
\end{figure*} 

\begin{table}
\begin{tabular}{|c|c|c|c|}\hline
~~n~~&SnIa data&BAO data&  H(z) data\\ \hline
\multirow{3}{*}{1}&$\chi^2_{min}=608.61$&$\chi^2_{min}=2.29$&$\chi^2_{min}=17.06$\\ 
&$w_0=-0.98$&$w_0=-1.0$&$w_0=-1.0$\\
&$\Omega_{m}=0.21$&$\Omega_{m}=0.28$&$\Omega_{m}=0.27$\\
&$\phi_{in}=1.7$&$\phi_{in}=10.0$&$\phi_{in}=3.1$\\ \hline
\multirow{3}{*}{2}&$\chi^2_{min}=608.59$&$\chi^2_{min}=2.41$&$\chi^2_{min}=17.06$\\ 
&$w_0=-0.98$&$w_0=-1.0$ &$w_0=-1.0$\\
&$\Omega_{m}=0.22$&$\Omega_{m}=0.28$&$\Omega_{m}=0.27$\\
&$\phi_{in}=6.2$&$\phi_{in}=10.0$&$\phi_{in}=5.8$\\ \hline
\multirow{3}{*}{3}&$\chi^2_{min}=608.54$&$\chi^2_{min}=2.65$&$\chi^2_{min}=17.05$\\ 
&$w_0=-0.98$&$w_0=-1.0$&$w_0=-1.0$\\
&$\Omega_{m}=0.21$&$\Omega_{m}=0.28$&$\Omega_{m}=0.27$\\
&$\phi_{in}=5.5$&$\phi_{in}=10.0$&$\phi_{in}=8.4$\\ \hline
\end{tabular}
  \caption{The above table shows the value of the parameters corresponding to minimum value of $\chi^2$ for the potential $V = M^{4-n}\phi^{n}$ .}
  \label{tab:thaw_pow_chi}
\end{table}

\begin{figure*}[t]
\centering
\begin{tabular}{ccc}
\includegraphics[scale=0.3,angle=270]{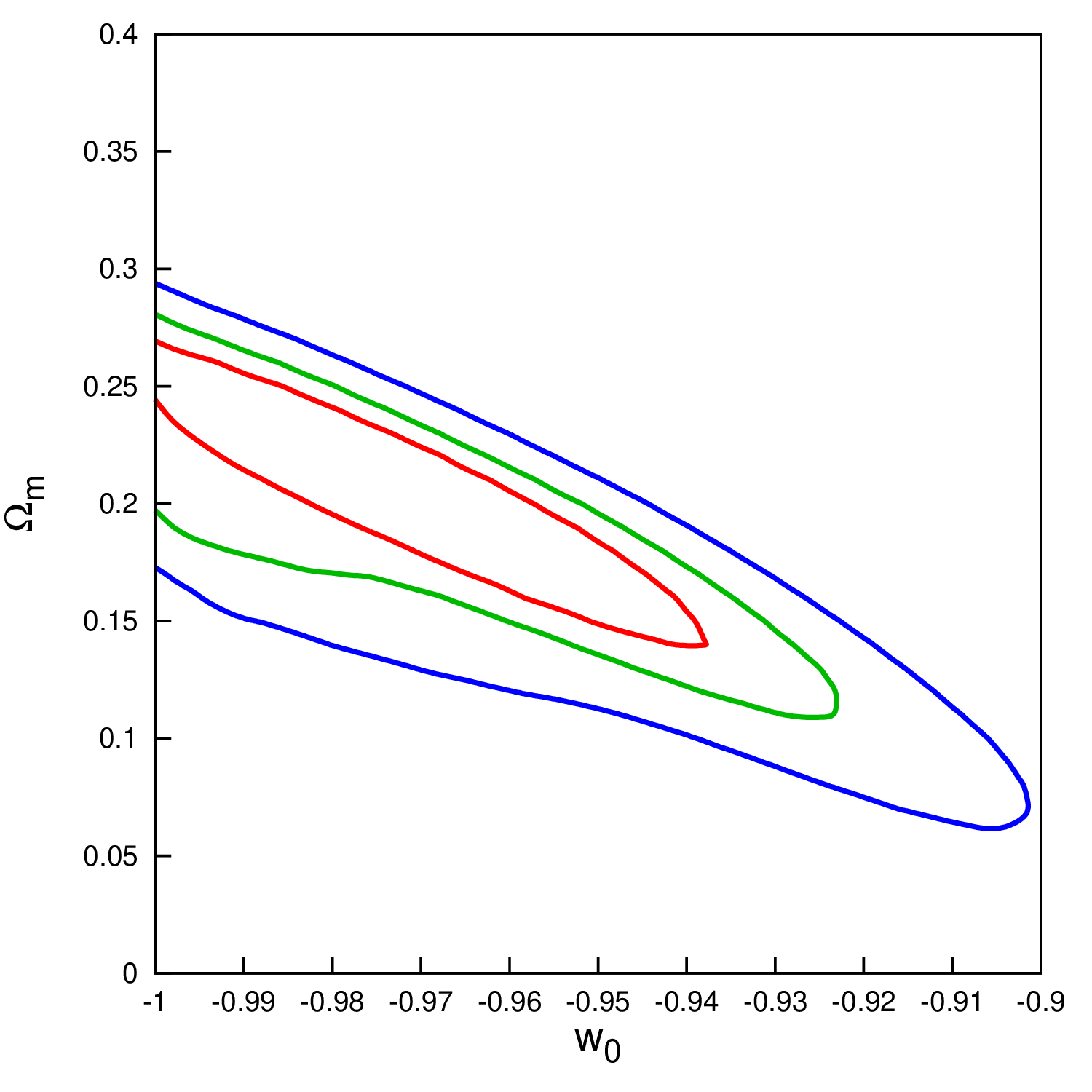}&\includegraphics[scale=0.3,angle=270]{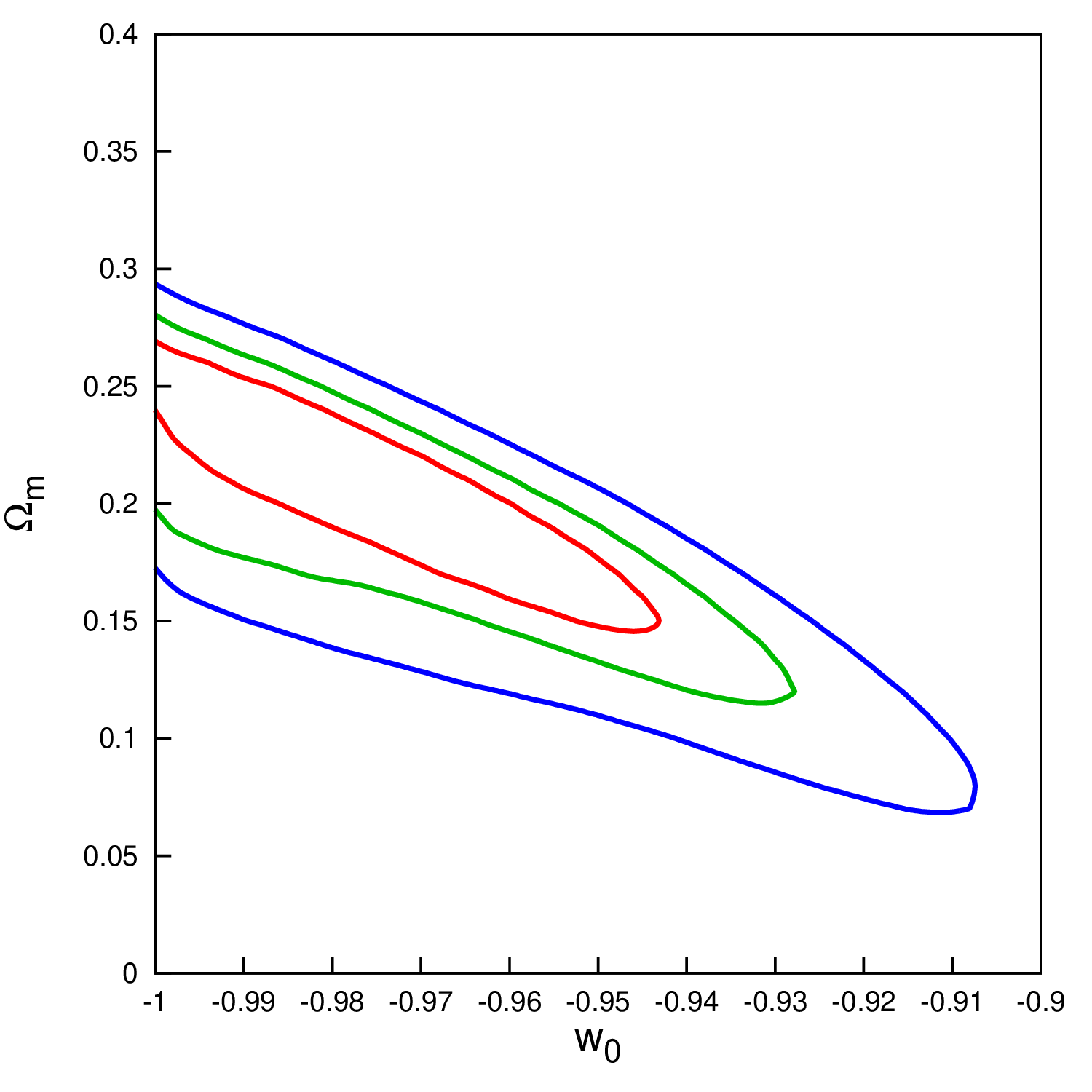}&\includegraphics[scale=0.3,angle=270]{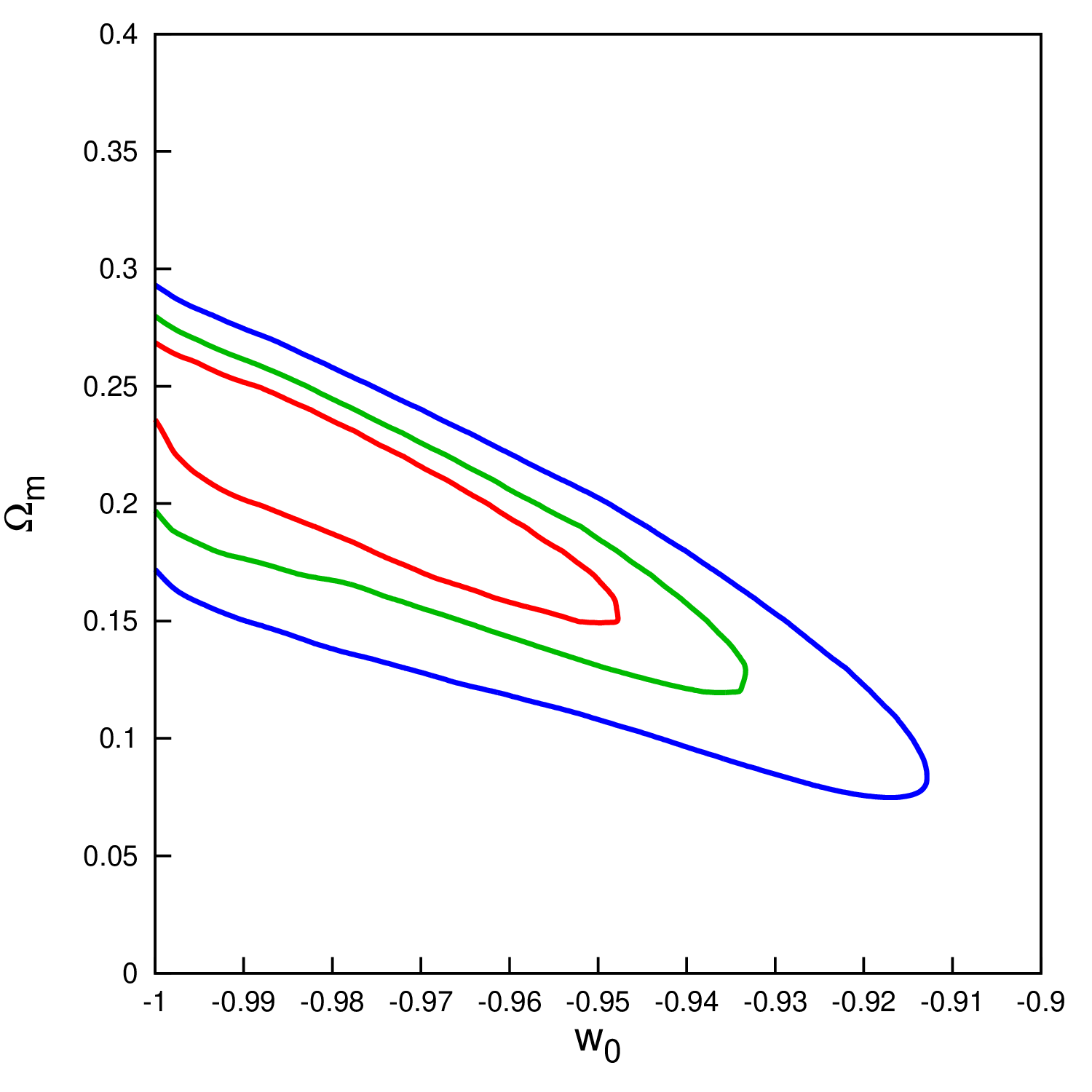}\\
\includegraphics[scale=0.3,angle=270]{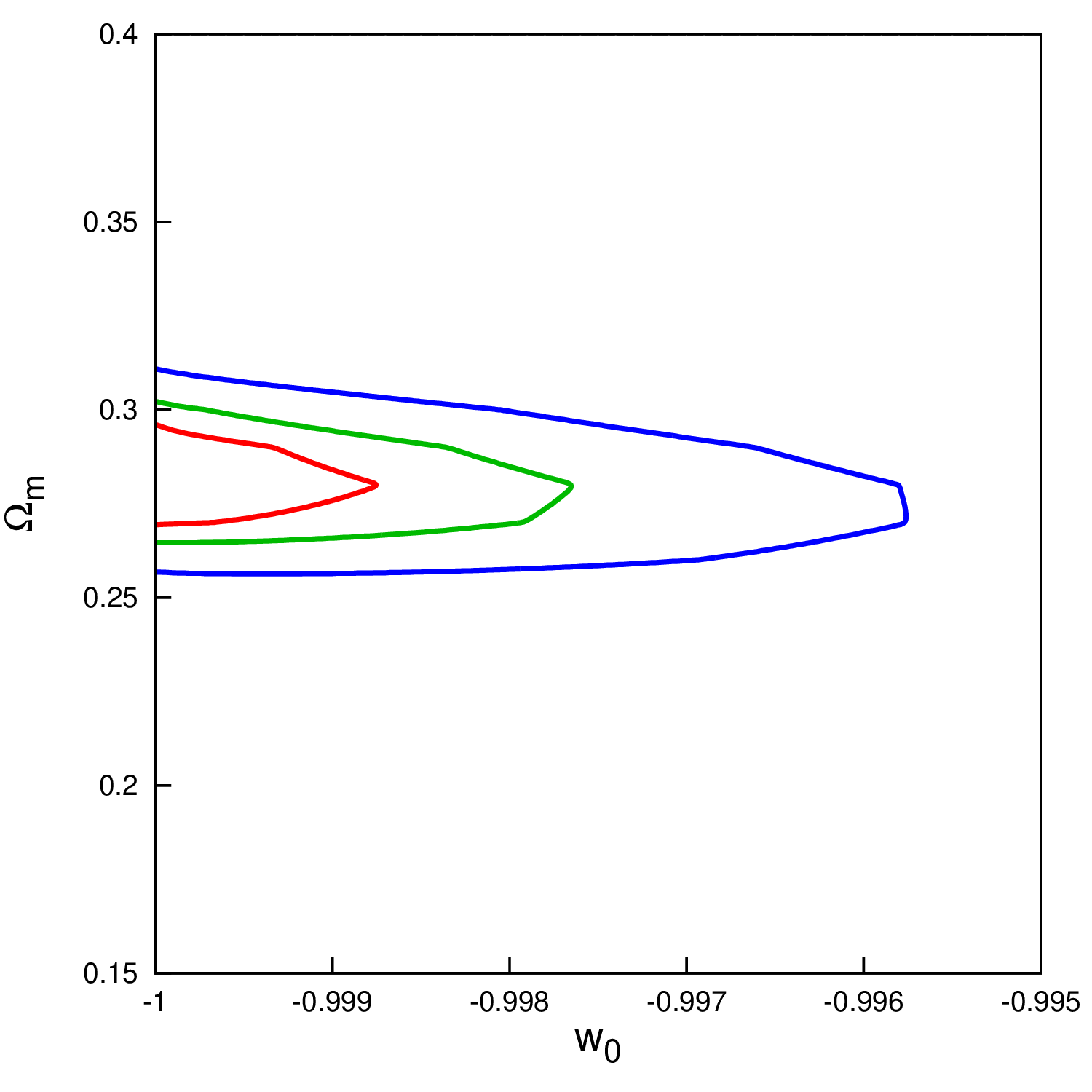}&\includegraphics[scale=0.3,angle=270]{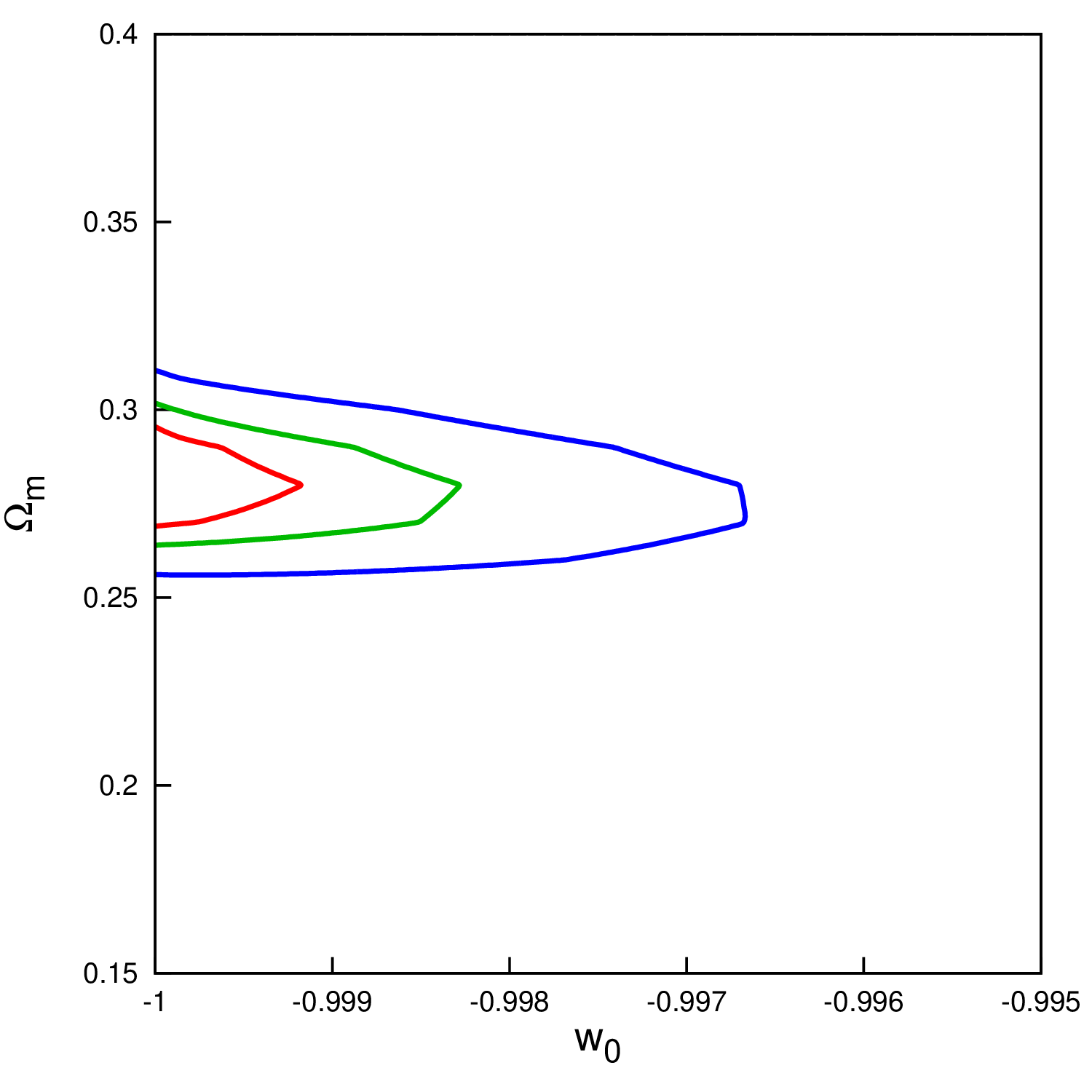}&\includegraphics[scale=0.3,angle=270]{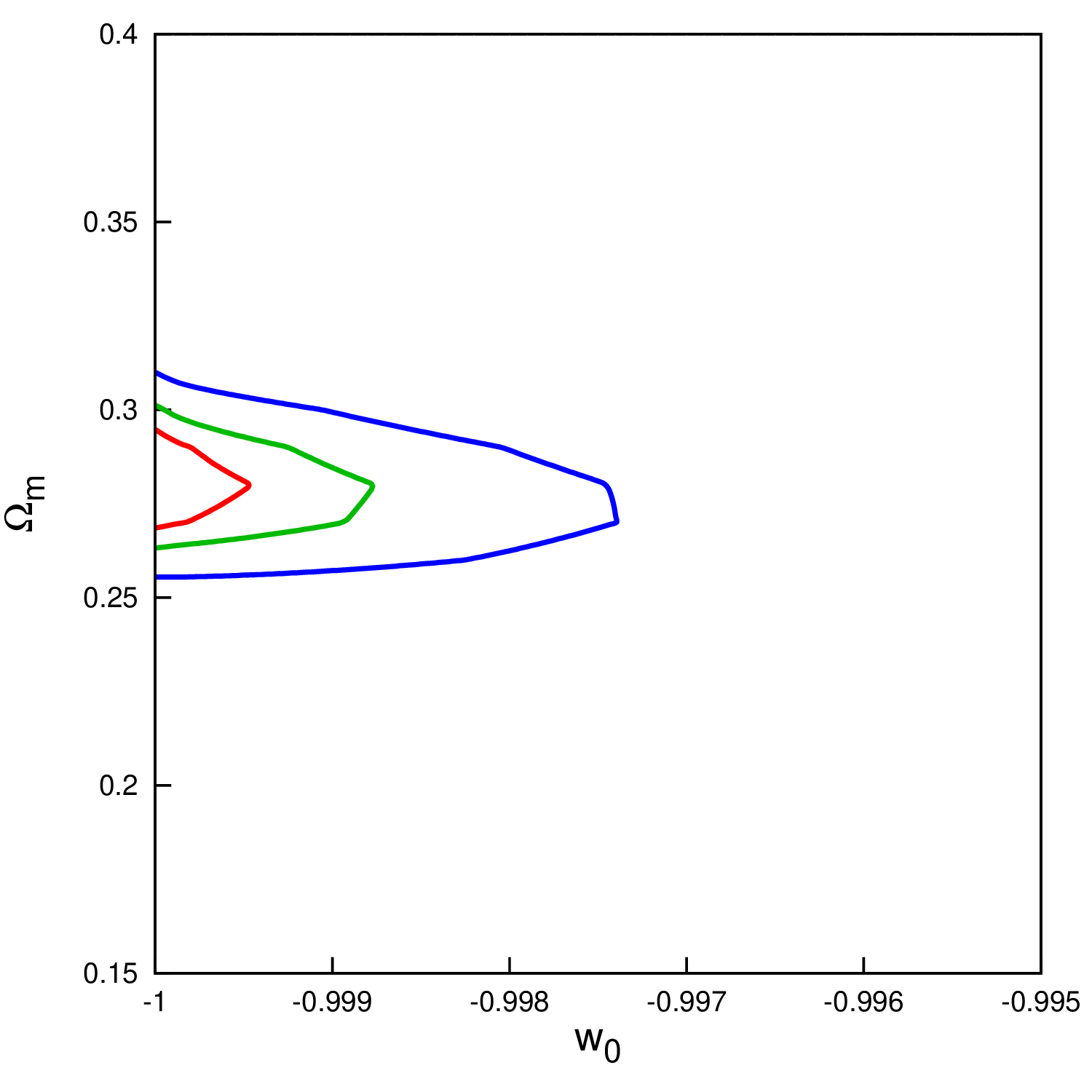}\\
\includegraphics[scale=0.3,angle=270]{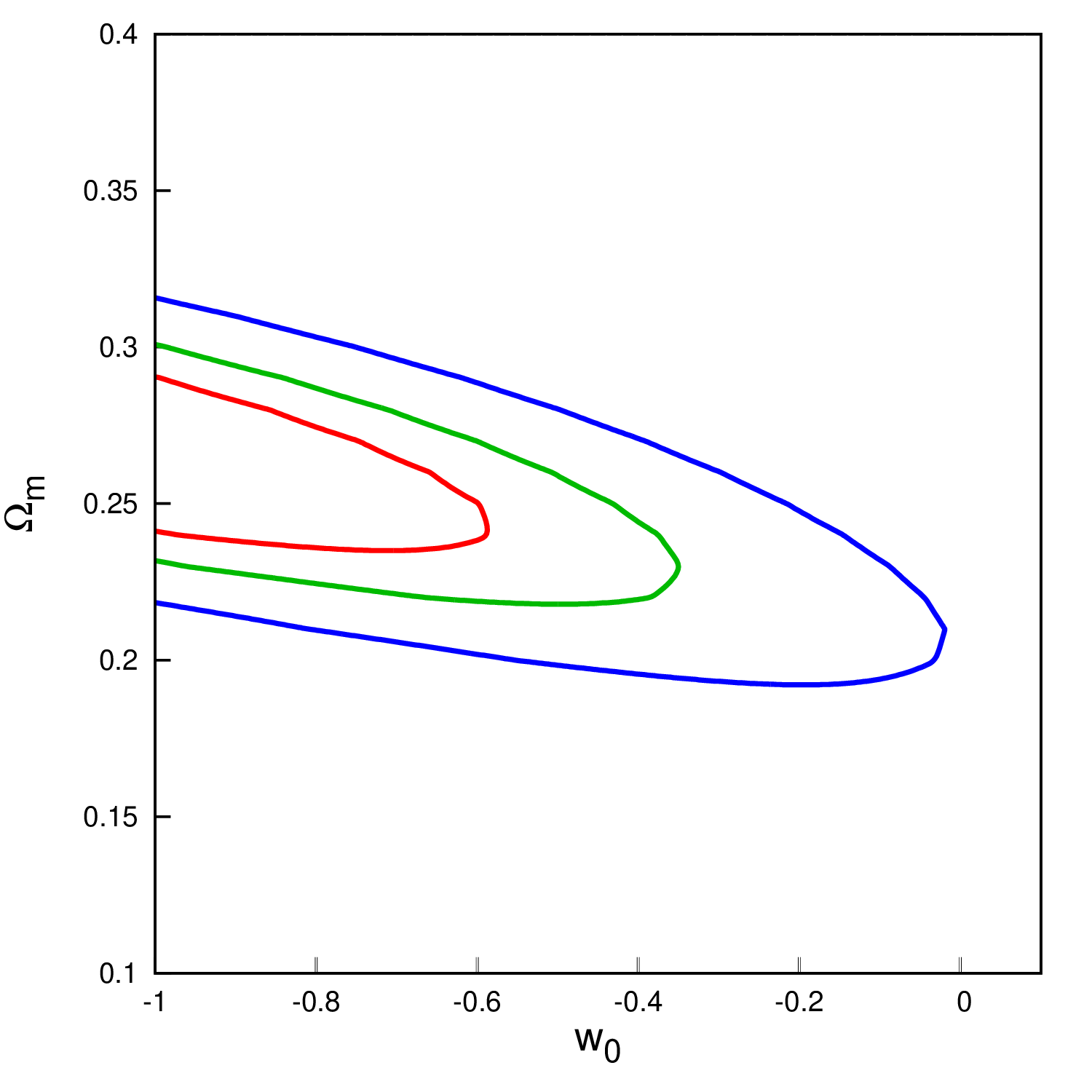}&\includegraphics[scale=0.3,angle=270]{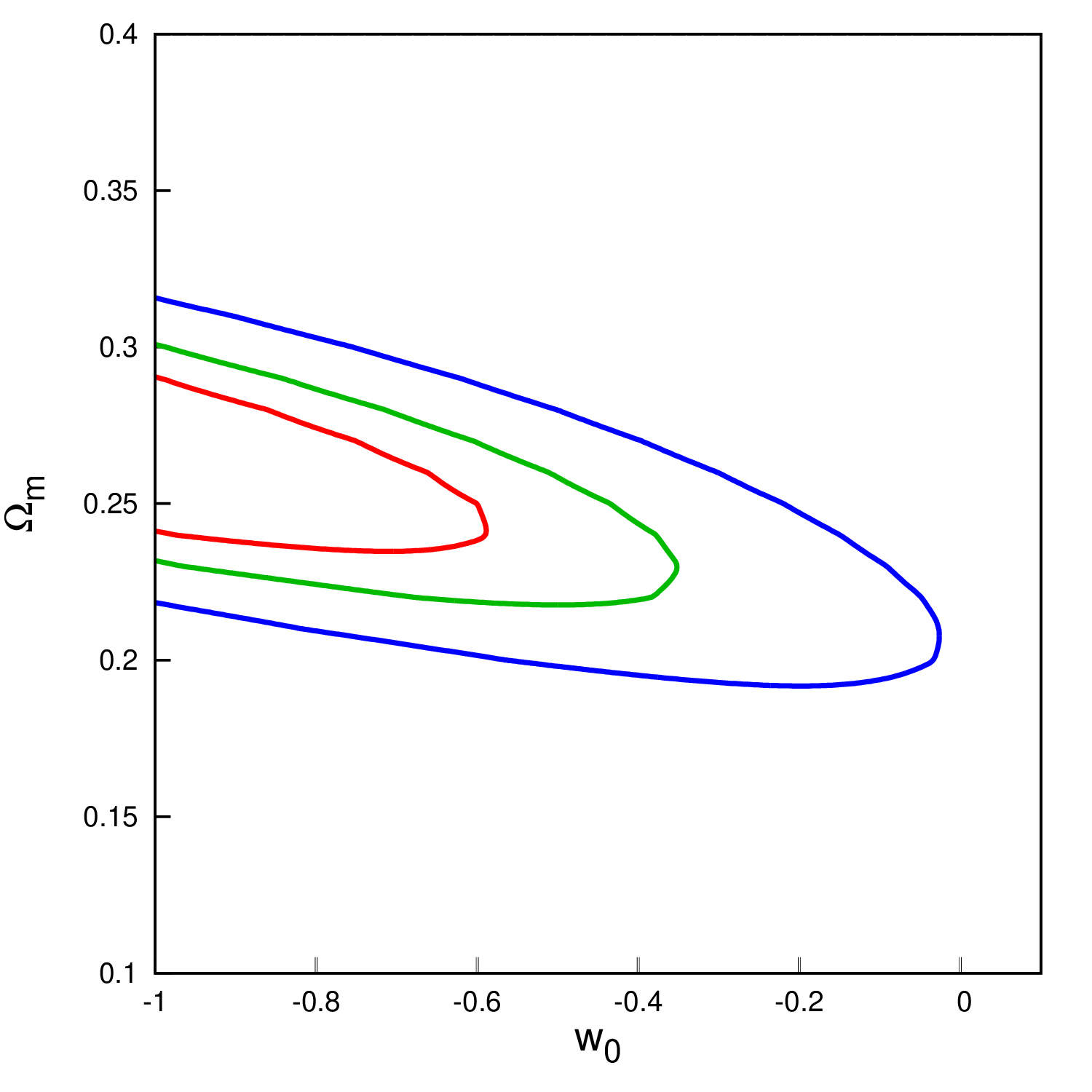}&\includegraphics[scale=0.3,angle=270]{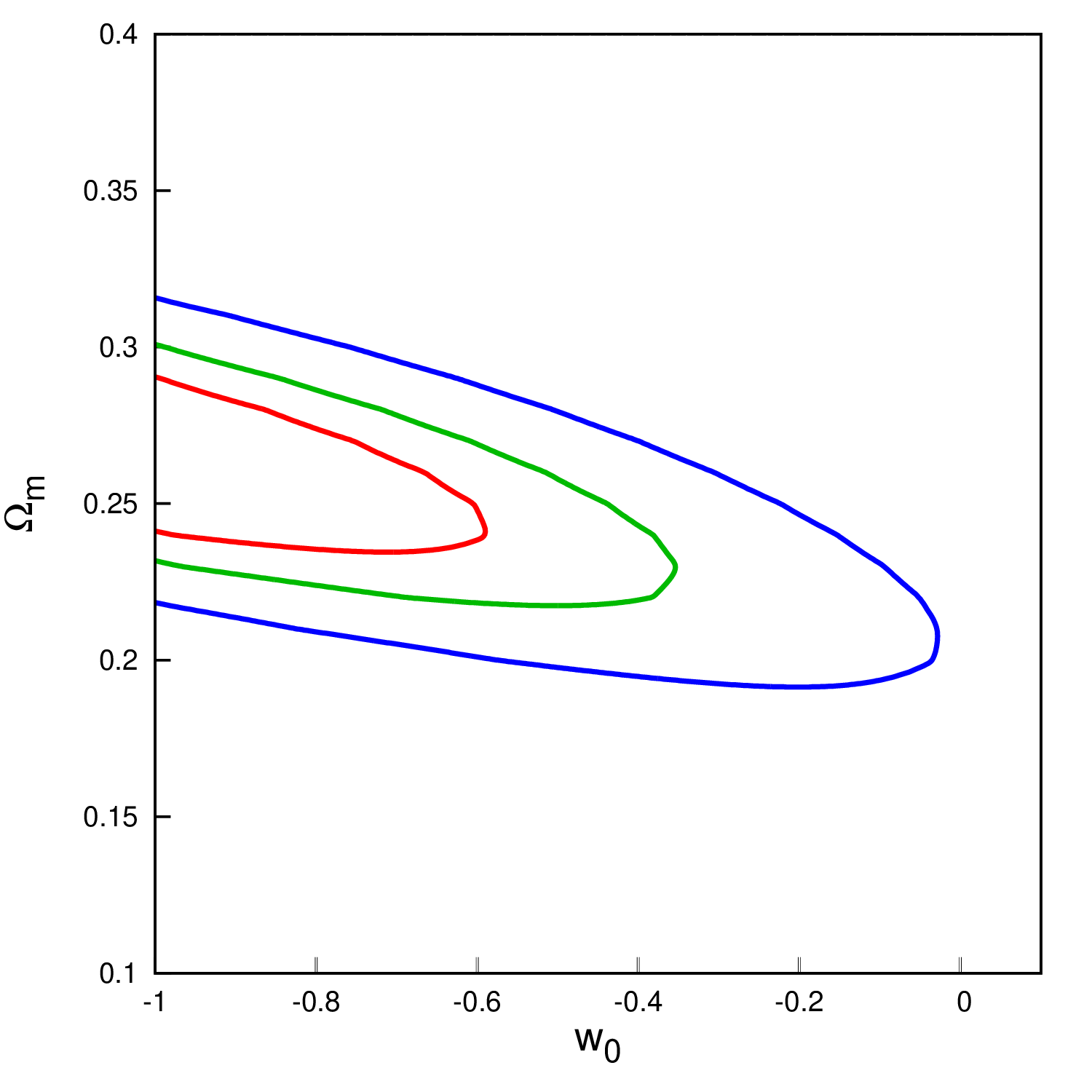}
\end{tabular}
\caption{The plots in this figure are  confidence  contours in $\Omega_{m}-w_0$ plane for the potential $V = V_0\phi^{n}$. In first row, from left to right, the plots are for SNIa data with  $n=1,2,3$ respectively.
  In the second row, the plots are for BAO data and the third row shows the plots for the H(z) data.} 
\label{fig::thaw_pow_data}
\end{figure*}

\begin{figure*}[t]
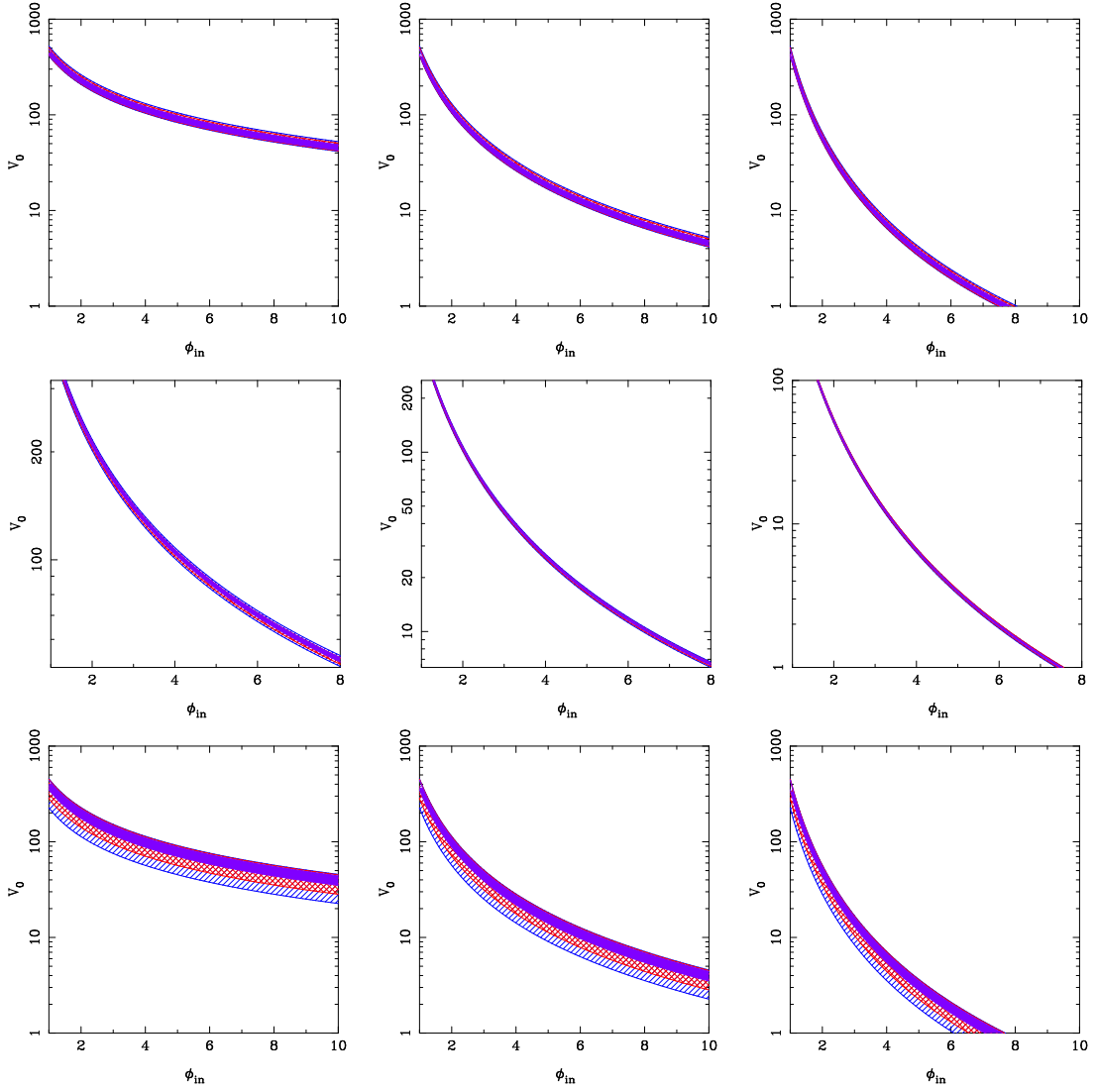

\centering
\begin{tabular}{ccc}
\includegraphics[scale=0.3]{fig7_a.ps}&\includegraphics[scale=0.3]{fig7_b.ps}&\includegraphics[scale=0.3]{fig7_c.ps}\\
\includegraphics[scale=0.3]{fig7_d.ps}&\includegraphics[scale=0.3]{fig7_e.ps}&\includegraphics[scale=0.3]{fig7_f.ps}\\
\includegraphics[scale=0.3]{fig7_g.ps}&\includegraphics[scale=0.3]{fig7_h.ps}&\includegraphics[scale=0.3]{fig7_i.ps}
\end{tabular}
\caption{The plots in the rows of the figure represent allowed region for $V_0=M^{4-n}$ corresponding to $1 \sigma$, $2 \sigma$ and 
 $3 \sigma$ confidence  region as a function of field $\phi_{in}$, for $V = V_0\phi^{n}$. Going from left to right, the first, second and
 third plots are obtained for potentials corresponding to $n$ equal to $1$, $2$ and $3$ respectively. In first row, the plots shows
 results for SNIa, the plots in second row are obtained for BAO data and the third row shows the plots for H(z) dataset. The solid blue region in the middle shows the allowed range of $V_0$ at $1 \sigma$ level, the red hatched region corresponds to $2 \sigma$ level and the region with blue slanted lines shows the $3 \sigma$ range. The white region is ruled out. 
} 
\label{fig::thawpow_v0phi}
\end{figure*}

\begin{table}
 \begin{tabular}{|c|l|l|l|}\hline
~~n~~&    SNIa data    &    BAO data    &    H(z) data    \\ \hline
&&&\\
1&-1.0$\leq$ $w_0$ $\leq$-0.92&-1.0$\leq$ $w_0$ $\leq$-0.995&-1.0$\leq$ $w_0$ $\leq$0.1\\ 
&0.1$\leq$ $\Omega_{m}$ $\leq$0.29&0.26$\leq$ $\Omega_{m}$ $\leq$0.31&0.19$\leq$ $\Omega_{m}$ $\leq$0.32\\
&1.0$\leq$ $\phi_{in}$ $\leq$10.0&1.0$\leq$ $\phi_{in}$ $\leq$ 10.0&1.0$\leq$ $\phi_{in}$ $\leq$ 10.0\\
&&&\\\hline
&&&\\
2&-1.0$\leq$ $w_0$ $\leq$-0.91&-1.0$\leq$ $w_0$ $\leq$-0.996&-1.0$\leq$ $w_0$ $\leq$0.1\\  
&0.1$\leq$ $\Omega_{m}$ $\leq$0.29&0.26$\leq$ $\Omega_{m}$ $\leq$0.31&0.18$\leq$ $\Omega_{m}$ $\leq$0.32\\
&1.0$\leq$ $\phi_{in}$ $\leq$10.0&1.9$\leq$ $\phi_{in}$ $\leq$ 10.0&1.0$\leq$ $\phi_{in}$ $\leq$ 10.0\\
&&&\\\hline
&&&\\
3&-1.0$\leq$ $w_0$ $\leq$-0.91&-1.0$\leq$ $w_0$ $\leq$-0.997&-1.0$\leq$ $w_0$ $\leq$0.08\\
&0.1$\leq$ $\Omega_{m}$ $\leq$0.29&0.26$\leq$ $\Omega_{m}$ $\leq$0.31&0.18$\leq$ $\Omega_{m}$ $\leq$0.32\\
&1.0$\leq$ $\phi_{in}$ $\leq$10.0&2.8$\leq$ $\phi_{in}$ $\leq$ 10.0&1.0$\leq$ $\phi_{in}$ $\leq$ 10.0\\ 
&&&\\\hline

 \end{tabular}
 \caption{The above table shows the 3$\sigma$ confidence limit for all the three data for the potential $V=V_0\phi^n$ for n=1,2,3.}
  \label{tab:thaw_pow_range}
\end{table}

In this section, we discuss the background cosmology and numerical
solutions for the different types of potentials that we have discussed
in the previous section.

\subsection{The exponential potential}
To study how the universe evolves in the presence of this potential, we
solve the Klein-Gordon equation, equation (\ref{kleingordon}), and Friedmann
equations for the scalar field, equation (\ref{field1}). 
In order to solve the equations, we define the following dimensionless
variable:
\begin{eqnarray}
\Phi= \frac{\phi}{M_{p}}.
\end{eqnarray}
The potential, then, takes the form
\begin{equation}
  V(\Phi) = M^{4} \exp({-\sqrt{\alpha} \Phi}). 
\end{equation}

In terms of the new variables, the cosmological equations can be written as 
\begin{eqnarray}
\ddot{\Phi} + 3\frac{\dot{a}}{a} \dot{\Phi} - \sqrt{\alpha} V_{0} \exp (-\sqrt{\alpha} \Phi) = 0,\\ \nonumber
\left(\frac{\dot{a}}{a}\right)^2 = H_0^2\frac{\Omega_{m}} {a^3} + \frac{\dot{\Phi}^2}{6} + \frac{V_0}{3} \exp (-\sqrt{\alpha} \Phi)
\end{eqnarray} 
where $V_{0}= \frac{M^4}{M_p^2}$ and $H_0$ is the present value of Hubble parameter.
For $\Omega_{total} = \Omega_{m} + \Omega_{\phi} = 1$, the initial
conditions are given by
\begin{equation}
V_{0} = \frac{3H_0^2}{2} (1 - \Omega_{m_i})(1 - w_{in})\exp(\sqrt{\alpha}\Phi_{in})
\label{equation::v0_thawexp}
\end{equation}

\begin{equation}
\dot{\Phi}_{in}^2=3H_0^2(1 - \Omega_{m_i})(1 + w_{in}); \hspace{0.5cm} \Phi_{in}=1.
\label{equation::phidot_thawexp}
\end{equation} 
The variables $\Omega_{m_i}$, $\Phi_{in}$ and $w_{in}$  are values of
non-relativistic matter density parameter, field and equation of state
parameter at some initial time $t=t_{i}$ (and $w_0$ represents the present day value of $w$).

By solving these coupled equations numerically, we solve for
 $\Phi$ and $\dot{\Phi}$ as  a function of the scale factor.
These values are, then, used to determine the value of equation of
state parameter $w$,  which in terms of the dimensionless parameters
is given by 
\begin{equation}
w = \frac{{\dot{\Phi}}^2 - 2V_{0}\exp(-\sqrt{\alpha} \Phi)}{{\dot{\Phi}}^2 + 2V_{0}\exp(-\sqrt{\alpha}\Phi)}.
\end{equation}
From the above equation we can see that, depending upon the form of potential
$V(\Phi)$, $w$ lies between $-1$ and  $+1$.

To study the evolution of the model, we evolve the system from early
time  to late time.  
We plot the results obtained for this potential in
figure~\ref{fig::thaw_exp_the}.
The  plot on the left in the first row shows the variation of $\Phi$ as a function
of scale factor (past to future).
The plot on the right shows the behaviour of equation of state parameter $w$ 
as scale factor  changes.
In the second row, the plot on left is for  energy density of the field
as a function of scale factor and the figure on the right  is the phase  plot
obtained for the model. 

\subsection {The Polynomial (concave) potential} 
The second potential of thawing class that we analysed is a power
potential given by 
\begin{equation}
V(\phi) = M^{4-n} \phi^n.
\end{equation}
The background equations then take the following form:
\begin{eqnarray}
\ddot{\phi} + 3\frac{\dot{a}}{a} \dot{\phi} + n V_{0} {\phi}^{n-1} = 0, \\ \nonumber
\left(\frac{\dot{a}}{a}\right)^2 = \frac{\Omega_{m}}{a^{3}} + \frac{\dot{\phi}^2}{6} + \frac{V_0 {\phi}^n}{3}.
\end{eqnarray} 
And equation of state becomes
\begin{equation}
w = \frac{{\dot{\phi}}^2 - 2V_{0} {\phi}^{n}}{{\dot{\phi}}^2 + 2V_{0} {\phi}^{n}}.
\end{equation}  
The value of $V_0$ for this potential is found to be   
\begin{equation}
V_{0} = \frac{3H_0^2}{16\pi G} (1 - \Omega_{m_i})(1 - w_{in})\phi^{-n}_{in}
\label{equation::v0_thawpow}
\end{equation}
and the initial value of field velocity, $\dot{\phi}_{in}$ is given by
\begin{equation}
\dot{\phi}_{in}=\pm\sqrt{\frac{3H_0^2}{8\pi G}(1 - \Omega_{m_i})(1 + w_{in})}.
\label{equation::phidot_thawpow}
\end{equation} 
As mentioned earlier, to study this potential, we consider $n=1,2,3$, corresponding to different background evolution.
 
We plot the results obtained by evolving the system from past to
present and then  to future for this potential in
figure~\ref{fig::thaw_the}. 
In the first row, the plot on left shows the variation of $\phi$ as a
function  of scale factor.
The next plot shows the behaviour of equation of state parameter $w$
for the potential with respect to scale factor, $a$.
In the second row, the plot on left is for energy density of the field
as a function of $a$ and the second figure is the phase  plot
obtained for the polynomial potential.

\section{Observational constraints on parameters}
\label{sec::results}

In this section, we discuss the results obtained by using the three
different cosmological observations in the analysis. 
For the data analysis we use the $\chi^2$ minimisation technique.
The observational data consists of $n$ points of
observables ($X_{n,ob}$), such as luminosity distance for supernova data or angular
diameter distance for the BAO data, at a particular redshift ($z_n$), along with 
error associated with the observable ($\sigma_n$). 
In this technique, we calculate the same observable
quantity ($X_{n,th}$) at the same redshift, with the equation
state parameter obtained by solving cosmological equations in the
presence of scalar field with a particular potential $V(\phi)$.  
The $\chi^2$ measures the goodness of fit i.e., by how much
the observational value differs in comparison to  theoretically expected
value and is defined as   
\begin{eqnarray}\label{chisq}
\chi^2 = \sum_k\left[\frac{X_{k,ob}-X_{k,th}}{\sigma_k}\right]^2
\end{eqnarray}
We have listed the priors used for the analysis in table \ref{table::priors}.

For the exponential
potential, $V(\phi)=M^4\exp(-\sqrt{\alpha}\phi/M_p)$, the free parameters  are the dark energy equation of state parameter $w_0$, 
matter density parameter $\Omega_m$ and $\alpha$ in combination with the present day value of $\phi/M_p$, where $M_p$ is the Plank mass. 
In figure \ref{fig::thaw_exp_data}, we show the $1 \sigma$, $2 \sigma$
and $3 \sigma$ confidence  contours in $\Omega_{m}-w_0$ plane.
Here, $\Omega_m$ and $w_0$ denote the present day value of non-relativistic matter density parameter and present day dark energy equation of state parameter.
The plot on the left is from SNIa data, the plot in the middle is for BAO data and the plot on the right shows the results from H(z) data. 
To obtain the contours we have marginalised over the entire range of the third parameter $\alpha$.
The minimum value of $\chi^2$ ($\chi^2_{min}$) and the constraints obtained for the
parameters are listed in table \ref{tab:thaw_exp_chi}.
BAO data provides the narrowest constraints on $\Omega_m$ and on the
upper limit of $w_0$; none of the data sets provide a lower limit on
$w_0$. 
The Hubble data constrains $\Omega_m$ strongly but it allows the
regions of $w_0$ within $3\sigma$ limits, which gives decelerated
expansion. 
Supernovae data allows the maximum range in $\Omega_m$; between
$\Omega_m=0.08$ to $0.31$ and the range of $w_0$ below $w\le-0.87$, and
it does not allow for a decelerated expansion of the universe. 

In the first row of figure~\ref{fig::thawexp_alphaphi}, we present the confidence
contours corresponding to $1\sigma$, $2\sigma$ and $3\sigma$ levels in
$V_0$ and $\sqrt{\alpha}\Phi$ plane. 
Here, we show the results for the range $0-1$ of $\sqrt{\alpha}\Phi$.
We find that the most stringent constraints are provided for BAO dataset, 
and the widest range is allowed for H(z) data and for SNIa dataset 
the range lies between the range provided by other two datasets.  
In figure \ref{fig::thawexp_alphaphi}, we show the allowed range of
$\sqrt{\alpha}\Phi$ and $w_0$ for different datasets in second row and in
third row we show the constraints on $\sqrt{\alpha}\Phi$ versus
$\Omega_m$.  
The first plot is obtained for SNIa, second plot is obtained for BAO
and third plot is the result from H(z) data respectively. 
The results are consistent with the confidence contours of
Fig. \ref{fig::thaw_exp_data}, this is because the value of $V_0$
depends upon both $\Omega_m$ and $w_0$.

We now discuss the results obtained for the exponential potential, $V(\phi)=M^{4-n}\phi^n$.
This gives us three different potentials, as $n$ takes three values; $n=1,2$ and $3$. 
The free parameters in the analysis for each of these potentials are
$w_0$, nonrelativistic matter density parameter $\Omega_m$ and the initial value of the field
$\phi_{in}$. 
The figure \ref{fig::thaw_pow_data} shows the $1 \sigma$, $2 \sigma$ and $3 \sigma$ confidence contours in $\Omega_{m}-w_0$ plane. %Here, $\Omega_m$ and $w_0$ denotes the present day value of non-relativistic matter density parameter and present day dark energy equation of state parameter.
The contours in the first row are obtained from analysis of SNIa
data, second row represents plots from BAO data and the third row
shows results for H(z) dataset. 
The contours in first, second and third columns are for $n$ $=$ $1$, $2$ and $3$ respectively.
The $2-D$ contours in $w_0-\Omega_m$ plane are obtained by marginalising over the third parameter $\phi_{in}$, the initial value of the field.
The value of the minimum  $\chi^2$ is listed in table
\ref{tab:thaw_pow_chi} and the constraints on the parameters are
listed in table \ref{tab:thaw_pow_range}.
Again, the most stringent constraints are provided by BAO data followed by SNIa and H(z) data.
The H(z) data also allows models with decelerated expansion for all three values of $n$, within $3\sigma$ limit.
We find that for a dataset, the tightest range is given for the
potential corresponding to $n=3$; as the value of $n$ goes from $3$ to
$1$, the allowed range for $w_0$ increases for all three datasets.  
None of the data sets provide a lower limit on $w_0$ and as the value of
$n$ increases the contours move towards $w=-1$, the cosmological
constant model. 
All the three datasets constrain $\Omega_m$ well, with
SNIa giving maximum allowed range for this parameter. 

In figure~\ref{fig::thawpow_v0phi}, we show the allowed range of
$V_0=M^{4-n}$ corresponding to $1\sigma$, $2\sigma$ and $3\sigma$
confidence regions as a function of field $\phi_{in}$. 
The scheme of plots is same as in figure~\ref{fig::thaw_pow_data}.
As the value of $\phi_{in}$ increases, the allowed range of values of $V_0$
decreases. 
This trend is same for the three datasets for all values of $n$. 
The maximum value $V_0$ is required for a smaller value of $\phi_{in}$. 
The maximum range is allowed by H(z) data and the narrowest range is
provided by BAO data for all values of $n$. 
The results are consistent with the confidence contours in plane
$w_0-\Omega_m$ of figure \ref{fig::thaw_pow_data} as the value of $V_0$
depends upon those parameters (see equation~(\ref{equation::v0_thawpow})). 
The solid blue region in the middle is the allowed $V_0$ region at $1\sigma$ 
level, the hatched lines (red) and the slanted lines (blue) regions represent 
$2\sigma$ and $3\sigma$ regions respectively.

\section{Summary}
\label{sec::conclusions}
In this paper, we present current constraints on
canonical scalar field models of dark energy.
Restricting ourselves to thawing scalar field models: we present these
results for an exponential potential and power law potentials with
different exponents.
To constrain the model parameters, we have taken three different
observational datasets into account: the high redshift supernova
observations, the baryon acoustic oscillations in galaxy clustering, and,
data from direct measurements of Hubble parameter at different epochs.
The observations considered here are sensitive to different combinations
of cosmological and dark energy parameters and together these allow for a
small range of scalar field parameters.
We consider two classes of models: the exponential potential and the power
law potential where we have considered three integer exponents for the
latter case.
The exponential potential has two parameters, while
after fixing the power exponent, the polynomial model reduces to a
one parameter potential.
These models have been and continue to be focus of different dark
energy studies and hence fitting these with later and different
observations is well motivated.

In all the models considered in this paper, the most stringent
constraints are due to the Baryon Acoustic Oscillation data.
While this dataset allows for a moderately large range in the equation
of state parameter, the allowed range of the value of the matter
density  parameter is very strongly limited and hence enables ruling out a
large range of parameters.
The supernova data restricts the parameter space with the allowed region
showing a degeneracy between the matter density
parameter and the equation of state parameter.
The Hubble parameter determination data allows for the largest range
in the equation of state parameter and also allows for
non-accelerating solutions.
The observations do not entirely rule out any of the models considered
here although the allowed range of parameters is narrow.
In general, the models which closely emulate the background evolution
of a cosmological constant at low redshifts are preferred by these
observations.
This result is consistent with constraints on fluid models of dark
energy and other studies on scalar field dark energy models.
While the datasets do limit the range of parameters, it is important
to confirm and further tighten the constraints with forthcoming
observations.
The constraints obtained from purely distance measurement observations
can be further used as priors for dark energy studies, especially in
studies of structure formation.

\section{Acknowledgments}
\label{sec::acknowledgements}
The numerical work in this paper was done on the High Performance
Computing facility at IISER Mohali.

\end{document}